\documentclass[12pt]{article}
\begin{document}
\newcommand{\beq}{\begin{equation}}
\newcommand{\eeq}{\end{equation}}
\newcommand{\beqa}{\begin{eqnarray}}
\newcommand{\eeqa}{\end{eqnarray}}
\newcommand{\beqar}{\begin{eqnarray*}}
\newcommand{\eeqar}{\end{eqnarray*}}
\newcommand{\al}{\alpha}
\newcommand{\be}{\beta}
\newcommand{\del}{\delta}
\newcommand{\D}{\Delta}
\newcommand{\eps}{\epsilon}
\newcommand{\ga}{\gamma}
\newcommand{\Ga}{\Gamma}
\newcommand{\ka}{\kappa}
\newcommand{\nn}{\nonumber}
\newcommand{\inn}{\!\cdot\!}
\newcommand{\h}{\eta}
\newcommand{\ii}{\iota}
\newcommand{\kk}{\varphi}
\newcommand\F{{}_3F_2}
\newcommand{\la}{\lambda}
\newcommand{\La}{\Lambda}
\newcommand{\na}{\prt}
\newcommand{\Om}{\Omega}
\newcommand{\om}{\omega}
\newcommand{\p}{\phi}
\newcommand{\sig}{\sigma}
\renewcommand{\t}{\theta}
\newcommand{\z}{\zeta}
\newcommand{\ssc}{\scriptscriptstyle}
\newcommand{\eg}{{\it e.g.,}\ }
\newcommand{\ie}{{\it i.e.,}\ }
\newcommand{\labell}[1]{\label{#1}} 
\newcommand{\reef}[1]{(\ref{#1})}
\newcommand\prt{\partial}
\newcommand\veps{\varepsilon}
\newcommand{\pol}{\varepsilon}
\newcommand\vp{\varphi}
\newcommand\ls{\ell_s}
\newcommand\cF{{\cal F}}
\newcommand\cA{{\cal A}}
\newcommand\cS{{\cal S}}
\newcommand\cT{{\cal T}}
\newcommand\cV{{\cal V}}
\newcommand\cL{{\cal L}}
\newcommand\cM{{\cal M}}
\newcommand\cN{{\cal N}}
\newcommand\cG{{\cal G}}
\newcommand\cH{{\cal H}}
\newcommand\cI{{\cal I}}
\newcommand\cJ{{\cal J}}
\newcommand\cl{{\iota}}
\newcommand\cP{{\cal P}}
\newcommand\cQ{{\cal Q}}
\newcommand\cg{{\it g}}
\newcommand\cR{{\cal R}}
\newcommand\cB{{\cal B}}
\newcommand\cO{{\cal O}}
\newcommand\tcO{{\tilde {{\cal O}}}}
\newcommand\bg{\bar{g}}
\newcommand\bb{\bar{b}}
\newcommand\bH{\bar{H}}
\newcommand\bX{\bar{X}}
\newcommand\bK{\bar{K}}
\newcommand\bR{\bar{R}}
\newcommand\bZ{\bar{Z}}
\newcommand\bxi{\bar{\xi}}
\newcommand\bphi{\bar{\phi}}
\newcommand\bpsi{\bar{\psi}}
\newcommand\bprt{\bar{\prt}}
\newcommand\bet{\bar{\eta}}
\newcommand\btau{\bar{\tau}}
\newcommand\hF{\hat{F}}
\newcommand\hA{\hat{A}}
\newcommand\hT{\hat{T}}
\newcommand\htau{\hat{\tau}}
\newcommand\hD{\hat{D}}
\newcommand\hf{\hat{f}}
\newcommand\hg{\hat{g}}
\newcommand\hp{\hat{\phi}}
\newcommand\hi{\hat{i}}
\newcommand\ha{\hat{a}}
\newcommand\hb{\hat{b}}
\newcommand\hQ{\hat{Q}}
\newcommand\hP{\hat{\Phi}}
\newcommand\hS{\hat{S}}
\newcommand\hX{\hat{X}}
\newcommand\tL{\tilde{\cal L}}
\newcommand\hL{\hat{\cal L}}
\newcommand\tG{{\widetilde G}}
\newcommand\tg{{\widetilde g}}
\newcommand\tphi{{\widetilde \phi}}
\newcommand\tPhi{{\widetilde \Phi}}
\newcommand\te{{\tilde e}}
\newcommand\tk{{\tilde k}}
\newcommand\tf{{\tilde f}}
\newcommand\ta{{\tilde a}}
\newcommand\tb{{\tilde b}}
\newcommand\tR{{\tilde R}}
\newcommand\teta{{\tilde \eta}}
\newcommand\tF{{\widetilde F}}
\newcommand\tK{{\widetilde K}}
\newcommand\tE{{\widetilde E}}
\newcommand\tpsi{{\tilde \psi}}
\newcommand\tX{{\widetilde X}}
\newcommand\tD{{\widetilde D}}
\newcommand\tO{{\widetilde O}}
\newcommand\tS{{\tilde S}}
\newcommand\tB{{\widetilde B}}
\newcommand\tA{{\widetilde A}}
\newcommand\tT{{\widetilde T}}
\newcommand\tC{{\widetilde C}}
\newcommand\tV{{\widetilde V}}
\newcommand\thF{{\widetilde {\hat {F}}}}
\newcommand\Tr{{\rm Tr}}
\newcommand\tr{{\rm tr}}
\newcommand\STr{{\rm STr}}
\newcommand\hR{\hat{R}}
\newcommand\M[2]{M^{#1}{}_{#2}}

\newcommand\bS{\textbf{ S}}
\newcommand\bI{\textbf{ I}}
\newcommand\bJ{\textbf{ J}}

\begin{titlepage}
\begin{center}

\vskip 2 cm
{\LARGE \bf    On NS-NS couplings   at order $\alpha'^3$
 }\\
\vskip 1.25 cm
   Mohammad R. Garousi\footnote{garousi@um.ac.ir}

\vskip 1 cm
{{\it Department of Physics, Faculty of Science, Ferdowsi University of Mashhad\\}{\it P.O. Box 1436, Mashhad, Iran}\\}
\vskip .1 cm
 \end{center}

\begin{abstract}
 Recently, imposing the gauge-symmetry and the T-duality constraints on the string frame effective actions of   type II superstring theories at order $\alpha'^3$, the NS-NS couplings   have been found in a particular scheme which has 445 couplings in 15 different structures.     In this paper, using a field redefinition, we write them  in terms of 251 couplings which appear in  9 different structures. The 9 structures involve only Riemann curvature, $H_{\alpha\beta\mu}$ and $\nabla_{\alpha}H_{\beta\mu\nu}$. The number of couplings in the  structures $R^4$, $H^8$, $H^6R$, $H^4R^2$, $H^2R^3$ are $2$, $2$, $1$, $7$, $22$, respectively, and the number of couplings in the structures $(\nabla H)^4$, $(\nabla H)^2R^2$, $(\nabla H)^2 H^4$, $(\nabla H)^2 H^2R$ are $12$, $22$, $77$, $106$, respectively. The new couplings are also fully consistent with  the sphere-level  S-matrix element of four NS-NS vertex operators.
  
\end{abstract}
\end{titlepage}


String theory is a candidate quantum theory for gravity   which includes  a finite number of massless fields and a  tower of infinite number of  massive fields reflecting the stringy nature of the gravity.
An efficient way to study different phenomena in this theory is to use an effective action  which includes     only  the massless fields \cite{Schwarz:1983qr,Schwarz:1983wa,Howe:1983sra,Carr:1986tk,Witten:1995ex}.
The effective action  has a double expansions. The genus-expansion which includes  the  classical tree-level  and a tower of  quantum  loop-level  corrections, and the   stringy-expansion which is an expansion in terms of  higher derivative couplings  at each loop level. 
It is shown in \cite{Veneziano:1991ek,Meissner:1991zj,Maharana:1992my,Meissner:1996sa,Kaloper:1997ux} that the tree-level effective action of bosonic string theory at two and four derivatives are invariant under T-duality  \cite{Giveon:1994fu,Alvarez:1994dn}. Using string field theory, it is proved in \cite{Sen:1991zi} that the full tree-level effective action of string theory should be  invariant under T-duality. Imposing this symmetry, the effective action at six-derivative level has been found in \cite{Garousi:2019mca}. 

The massless fields in  type II superstring theories are bosons and fermions. Using supersymmetry, their tree-level effective actions at two-derivative level have been found long  ago \cite{Schwarz:1983qr,Schwarz:1983wa,Howe:1983sra,Carr:1986tk}. The bosonic couplings in democratic form \cite{Fukuma:1999jt} in the string frame are
\beqa
 \quad \bS_0= -\frac{2}{\kappa^2}\int d^{10} x\sqrt{-g} \Big[e^{-2\Phi} \left(  R + 4\nabla_{a}\Phi \nabla^{a}\Phi-\frac{1}{12}H^2\right)+ \sum_{n=1}^{9}\frac{1}{n!}F^{(n)}\cdot F^{(n)}\Big]\labell{S0bf}
\eeqa
where $H$ is field strength of the $B$-field and $F^{(n)}$ is the R-R field strength. This action is  invariant under the T-duality \cite{Maharana:1992my,Fukuma:1999jt,Garousi:2019jbq}. The first stringy-correction to this action is at eight derivative order. It has been shown in \cite{Garousi:2020mqn} that the gauge symmetry alone requires to have  at least 872 independent NS-NS couplings. All other couplings can be converted to these couplings by the total derivative terms, by field redefinitions and by the  Bianchi identities.   Imposing the T-duality constraint on these 872 couplings, the NS-NS couplings in type II superstring theories have been found in the string frame   to be \cite{Garousi:2020gio}
\beqa
 \quad \bS_3= -\frac{2\alpha'^3 c}{\kappa^2}\int d^{10} x\sqrt{-g} \Big[e^{-2\Phi} \cL_3(g,B,\Phi)+ \cdots\Big]\labell{S3bf}
\eeqa
where dots represent the R-R and fermion fields in which we are not interested, and $c$ is an overall factor which can not be fixed by the T-duality constraint.  There are 445 non-zero NS-NS couplings in $\cL_3(g,B,\Phi)$ which appear in 15 different structures. There are only two couplings when there is no B-field, \ie 
\beqa
{\cal L}_3(g,\Phi)&=& 2 R_{\alpha}{}^{\epsilon}{}_{\gamma}{}^{\varepsilon} R^{\alpha \beta \gamma \delta} R_{\beta}{}^{\mu}{}_{\epsilon}{}^{\zeta} R_{\delta \zeta \varepsilon \mu} + R_{\alpha \beta}{}^{\epsilon \varepsilon} R^{\alpha \beta \gamma \delta} R_{\gamma}{}^{\mu}{}_{\epsilon}{}^{\zeta} R_{\delta \zeta \varepsilon \mu}\labell{RRf}
\eeqa  
The couplings  in this structure are exactly the couplings that have been found by the S-matrix and sigma-model calculations \cite{Gross:1986mw,Myers:1987qx,Grisaru:1986vi,Freeman:1986zh,Policastro:2008hg,Paban:1998ea,Garousi:2013zca} provided that   one chooses the overall parameter to be $c=-\z(3)/2^6$. All other 443 couplings appear in 14 different structures which involve $H$. We refer the interested reader to \cite{Garousi:2020gio} for the explicit form of these couplings.

 The  number  of structures and the number of couplings in each structure, in general,  are not unique. They   are changed under higher-derivative field redefinitions and under adding total derivative terms, \ie they are scheme dependent \cite{Metsaev:1987zx}. Except the couplings in \reef{RRf}, all other couplings in $\cL_3(g,B,\Phi)$ are changed under adding total derivative terms and under using  the field redefinition. Hence, it raises question about the number of structures and the number of couplings in a specific scheme. In this paper, we are going to show that the couplings found in \cite{Garousi:2020gio} can be written in another scheme in which there are  only 9 structures. The dilaton appears in this scheme only as the overall factor $e^{-2\Phi}$. Moreover, the number of couplings in this scheme is 251 which is  much less than the number of couplings found in \cite{Garousi:2020gio}.  

To add arbitrary  total derivative terms to $\cL_3(g,B,\Phi)$, we consider the most general total derivative terms at order $\alpha'^3$ in the string frame which have the following structure:
\beqa
\frac{\alpha'^3}{\kappa^2}\int d^{10}x \sqrt{-g}e^{-2\Phi} \mathcal{J}_3=\frac{\alpha'^3}{\kappa^2}\int d^{10}x\sqrt{-g} \nabla_\alpha (e^{-2\Phi}{\cal I}_3^\alpha) \labell{J3}
\eeqa
where the vector ${\cal I}_3^\alpha$ is   all possible  covariant and gauge invariant  terms at seven-derivative level with even parity.  Using the package   "xAct" \cite{Nutma:2013zea}, one finds 11941 covariant terms, \ie 
\beqa
{\cal I}_3^\alpha= &J_1 H^{\gamma\delta\epsilon}R^{\alpha\beta}R_{\beta\varepsilon\epsilon\theta}\nabla_{\delta}H_{\gamma}{}^{\varepsilon\theta}+\cdots
\eeqa
where the coefficients  $J_1,\cdots, J_{11941}$ are 11941 arbitrary parameters. 

The couplings in  $\cL_3(G,B,\Phi)$ are also in a  fixed field variables. One can change the field variables as 
\begin{eqnarray}
g_{\mu\nu}&\rightarrow &g_{\mu\nu}+\alpha'^3 \delta g^{(3)}_{\mu\nu}\nn\\
B_{\mu\nu}&\rightarrow &B_{\mu\nu}+ \alpha'^3\delta B^{(3)}_{\mu\nu}\nn\\
\Phi &\rightarrow &\Phi+ \alpha'^3\delta\Phi^{(3)}\labell{gbp}
\end{eqnarray}
where the tensors $\delta g^{(3)}_{\mu\nu}$, $\delta B^{(3)}_{\mu\nu}$ and $\delta\Phi^{(3)}$ are all possible covariant and gauge invariant terms at 6-derivative level.  $\delta g^{(3)}_{\mu\nu}$, $\delta\Phi^{(3)}$ contain even-parity terms and $\delta B^{(3)}_{\mu\nu}$ contains odd-parity terms \ie,
\beqa
 \delta g^{(3)}_{\alpha\beta}&=& g_1 H_{\{\alpha}{}^{\gamma\delta}H_{\beta\}\gamma}{}^{\epsilon}H_{\delta}{}^{\varepsilon\theta}H_{\epsilon\varepsilon}{}^{\eta}H_{\theta}{}^{\mu\nu}H_{\eta\mu\nu}+\cdots \nn\\
 \delta B^{(3)}_{\alpha\beta}&=& e_1 R^{\gamma\delta}R_{\delta\epsilon\varepsilon [\alpha}\nabla_{\beta ]}H_{\gamma}{}^{\epsilon\varepsilon}+\cdots \nn\\
 \delta\Phi^{(3)}&=&f_1 H_{\alpha}{}^{\delta\epsilon}R^{\beta\gamma}\nabla^{\alpha}\Phi\nabla_{\gamma}H_{\beta\delta\epsilon}+\cdots\labell{eq.12}
\eeqa
The coefficients $g_1,\cdots, g_{3440}$, $e_1,\cdots, e_{2843}$ and $f_1,\cdots, f_{705}$ are arbitrary parameters. When the field variables in $\!\!\bS_3$  are changed according to the above  field redefinitions, they produce some couplings at orders $\alpha'^6$ and higher in which we are not interested in this paper. However, when the NS-NS field variables in $\!\!\bS_0$  are changed,  up to some total derivative terms, the following   couplings  at order $\alpha'^3$ are produced:
\beqa
\delta \!\!\bS_0&\!\!\!\!\!=\!\!\!\!\!\!&\frac{\delta \!\!\bS_0}{\delta g_{\alpha\beta}}\delta g^{(3)}_{\alpha\beta}+\frac{\delta \!\!\bS_0}{\delta B_{\alpha\beta}}\delta B^{(3)}_{\alpha\beta}+\frac{\delta \!\!\bS_0}{\delta \Phi}\delta \Phi^{(3)}
= -\frac{2\alpha'^3}{\kappa^2}\int d^{10} x\sqrt{-g}e^{-2\Phi}\Big[\nn\\&&(\frac{1}{2} \nabla_{\gamma}H^{\alpha \beta \gamma} -  H^{\alpha \beta}{}_{\gamma} \nabla^{\gamma}\Phi)\delta B^{(3)}_{\alpha\beta} -(  R^{\alpha \beta}-\frac{1}{4} H^{\alpha \gamma \delta} H^{\beta}{}_{\gamma \delta}+ 2 \nabla^{\beta}\nabla^{\alpha}\Phi)\delta g^{(3)}_{\alpha\beta}
\nn\\
&&-2( R -\frac{1}{12} H_{\alpha \beta \gamma} H^{\alpha \beta \gamma} + 4 \nabla_{\alpha}\nabla^{\alpha}\Phi -4 \nabla_{\alpha}\Phi \nabla^{\alpha}\Phi)(\delta\Phi^{(3)}-\frac{1}{4}\delta g^{(3)\mu}{}_\mu) \Big]\nn\\
&\equiv &\frac{\alpha'^3}{\kappa^2}\int d^{10}x\sqrt{-g}e^{-2\Phi}\mathcal{K}_3
\eeqa
 Adding the total derivative terms and the field redefinition terms  to the action \reef{S3bf}, one finds new  action $S_3$, \ie
 \beqa
 S_3= -\frac{2\alpha'^3 c}{\kappa^2}\int d^{10} x\sqrt{-g} \Big[e^{-2\Phi} L_3(g,B,\Phi)+ \cdots\Big]\labell{S3bf1}
\eeqa
where the  Lagrangian  $L_3(G,B,\Phi)$ is related to the Lagrangian $\cL_3(G,B,\Phi)$ as   
\beqa
L_3&=&\cL_3+{\cal J}_3+{\cal K}_3\labell{DLK}
\eeqa
The action $\bS_3$ and $S_3$ are physically equivalent. There is no free parameter in $\cL_3(G,B,\Phi)$. Choosing different values for the arbitrary parameters in ${\cal J}_3$, ${\cal K}_3$, one would find different forms of couplings for the Lagrangian $L_3$.  Alternatively, if one chooses a specific form for the Lagrangian $L_3$ and the above equation has a solution for the arbitrary parameters in ${\cal J}_3$, ${\cal K}_3$, then that Lagrangian would be  physically the same as $\cL_3$. Otherwise that Lagrangian would  not correspond to an effective action of the superstring theory.   

To check that the above equation has solution, however,  one should write \reef{DLK}  in terms of independent couplings, \ie    one has to impose the following Bianchi identities:
\beqa
 R_{\alpha[\beta\gamma\delta]}&=&0\nn\\
 \nabla_{[\mu}R_{\alpha\beta]\gamma\delta}&=&0\labell{bian}\\
\nabla_{[\mu}H_{\alpha\beta\gamma]}&=&0\nn\\
{[}\nabla,\nabla{]}\mathcal{O}-R\mathcal{O}&=&0\nn
\eeqa
 To impose these Bianchi identities in gauge invariant form, one may contract the  left-hand side of each Bianchi identity with the NS-NS field strengths and their derivatives  to produce terms at order $\alpha'^3$. The coefficients of these terms are arbitrary. Adding these terms to the equation \reef{DLK}, then one can check whether or not it has solution.  Alternatively, to impose the  Bianchi identities in non-gauge invariant form, one may rewrite the terms in \reef{DLK} in  the local frame in which the first derivative of metric is zero, and  rewrite the terms in \reef{DLK} which have derivatives of $H$ in terms of B-field, \ie $H=dB$.  In this way,  the Bianchi identities satisfy automatically \cite{Garousi:2019cdn}.  This latter approach is easier to impose the Bianchi identities by computer. Moreover, in this approach one does not need to introduce another  large number of arbitrary parameters to include the Bianchi identities to the equation \reef{DLK}.  
 
We are looking for a Lagrangian which has minimum number of structures. The Lagrangian $\cL_3$ has 15 structures. Each structure has no term with more than two derivatives, \eg there is no coupling which has $\nabla\nabla H$.  In some of the structures the dilaton appears non-trivially, \eg there are couplings which has $\nabla\nabla\Phi$ (see \cite{Garousi:2020gio} for the explicit form of the couplings). To reduce the number of structures, we guess that there might be scheme in which there is no term with more than two derivatives as in $\cL_3$, and the dilaton appears in them trivially as the overall factor $e^{-2\Phi}$. To check this proposal, we write all contractions of Riemann curvature, $H$  and $\nabla H$, in which derivative is not contracted with $H$, \ie no  coupling  has term $\nabla_{\mu}H^{\mu\nu\alpha}$. Using the package   "xAct" \cite{Nutma:2013zea}, one finds there are 1141 such couplings in 9 different structures, \ie 
\beqa
 L_3=& c_1 H_{\alpha}{}^{\delta \epsilon} H^{\alpha \beta \gamma} \
H_{\beta}{}^{\varepsilon \mu} H_{\gamma}{}^{\zeta \eta} H_{\
\delta \varepsilon}{}^{\theta} H_{\epsilon \zeta}{}^{\iota} \
H_{\mu \iota}{}^{\kappa} H_{\eta \theta \kappa}+\cdots\labell{L3}
\eeqa
where $c_1,\cdots, c_{1141}$ are some parameters. When we have replaced this Lagrangian on the right-hand side of \reef{DLK} and rewrite all terms in \reef{DLK} in terms of independent non-gauge invariant form, we find that there is solution for the parameters. It means it is possible to rewrite the couplings in $\cL_3$ in the above 9 structures in $L_3$.  However, not all 1141 parameters are fixed in solving \reef{DLK}. It means we are still free to choose some of the parameters in $L_3$ to be zero. We have checked that  it is impossible to remove any more any of   the 9 structures in $L_3$.  

To find the minimum number of couplings in these 9 structures,  we choose each term in any structure, say $R^4$,  and check if it is zero, does the equation \reef{DLK} have solution or not? We keep only  the coefficients that the equation \reef{DLK} has no solution if they are zero,  and set all other coefficients in that structure  to be zero.  Then  we solve the equation \reef{DLK}  to find the non-zero coefficients.  In this way we have found that, as expected, the couplings in structure $R^4$ are those in \reef{RRf}, \ie $L_3^{R^4}=\cL_3(g,\Phi)$. 

To find the minimum number of couplings with structure $H^8$, we  again keep only the terms that if one sets  any of them to zero, the equation \reef{DLK} has no solution. In this case there are two terms. Setting all parameters in the structure $H^8$  to zero, except those two parameters, and solving \reef{DLK}, one finds the following couplings in the structure $H^8$:
\beqa
L_3^{H^8}&=&\frac{1}{48} H_{\alpha}{}^{\delta \epsilon} H^{\alpha \beta 
\gamma} H_{\beta}{}^{\varepsilon \mu} H_{\gamma}{}^{\zeta 
\eta} H_{\delta \varepsilon}{}^{\theta} H_{\epsilon 
\zeta}{}^{\iota} H_{\theta \iota \kappa} H_{\mu 
\eta}{}^{\kappa}\nn\\&& -  \frac{9}{128} H_{\alpha}{}^{\delta 
\epsilon} H^{\alpha \beta \gamma} H_{\beta}{}^{\varepsilon 
\mu} H_{\gamma}{}^{\zeta \eta} H_{\delta 
\varepsilon}{}^{\theta} H_{\epsilon \zeta}{}^{\iota} H_{\eta 
\theta \kappa} H_{\mu \iota}{}^{\kappa}\labell{H8}
\eeqa 
In fact, it has been observed in \cite{Garousi:2020mqn} that the coefficients of five terms in structure $H^8$ are invariant under field redefinition. The T-duality fix three of them to be zero and the other two are as above. However, in other schemes, there are the above two couplings as well as some other couplings, \eg there are  five couplings in the scheme used in \cite{Garousi:2020gio}.

Having set all parameters in the structure $H^8$ to zero, except the above two terms, we now find the minimum number of couplings with   structure $RH^6$. To this end, we  again keep only the terms that if one set to zero any of them, the equation \reef{DLK} has no solution. In this case there is only one term. Setting all parameters in the structure $RH^6$  to zero, except that parameter, and solving \reef{DLK}, one finds the following coupling in the structure $RH^6$:
\beqa
L_3^{RH^6}&=&\frac{9}{8} H_{\alpha}{}^{\delta \epsilon} H^{\alpha \beta \gamma} 
H_{\beta}{}^{\varepsilon \mu} H_{\gamma}{}^{\zeta \eta} H_{\delta 
\varepsilon}{}^{\theta} H_{\epsilon \zeta}{}^{\iota} R_{\mu \iota \eta \theta}\labell{RH6}
\eeqa
It is interesting to note there are 25 couplings  in the structure $RH^6$ in the scheme used in \cite{Garousi:2020gio}, whereas in the present scheme there is only one term. 

Having set all parameters in the structures $H^8$ and $RH^6$ to  zero, except the couplings in \reef{H8} and \reef{RH6}, we now find the minimum number of couplings   with structure $R^2H^4$. To this end, we  again keep only the terms that if one set to zero any of them, the equation \reef{DLK} has no solution. In this case there are 7 terms. Setting all parameters in the structure $R^2H^4$  to zero, except those 7 parameters, and solving \reef{DLK}, one finds the following couplings in the structure $R^2H^4$:
\beqa
L_3^{R^2H^4}&=&\frac{7}{2} H_{\alpha}{}^{\delta \epsilon} H^{\alpha \beta 
\gamma} H_{\beta}{}^{\varepsilon \mu} H_{\delta}{}^{\zeta 
\eta} R_{\gamma}{}^{\theta}{}_{\varepsilon 
\zeta} R_{\epsilon \theta \mu \eta} -  
\frac{11}{4} H_{\alpha}{}^{\delta \epsilon} H^{\alpha \beta 
\gamma} H_{\beta}{}^{\varepsilon \mu} H^{\zeta \eta \theta} R_{\gamma \zeta \delta \epsilon} R_{\varepsilon \eta \mu \theta}\nn\\&& -  \frac{5}{2} 
H_{\alpha}{}^{\delta \epsilon} H^{\alpha \beta \gamma} 
H_{\beta \delta}{}^{\varepsilon} H_{\gamma}{}^{\mu \zeta} 
R_{\epsilon}{}^{\eta}{}_{\varepsilon}{}^{\theta} 
R_{\mu \eta \zeta \theta} -  \frac{1}{2} 
H_{\alpha}{}^{\delta \epsilon} H^{\alpha \beta \gamma} 
H_{\beta \delta}{}^{\varepsilon} H_{\gamma \epsilon}{}^{\mu} R_{\varepsilon}{}^{\zeta \eta \theta} 
R_{\mu \eta \zeta \theta}\nn\\&& - 2 
H_{\alpha}{}^{\delta \epsilon} H^{\alpha \beta \gamma} 
H_{\beta}{}^{\varepsilon \mu} H_{\delta}{}^{\zeta \eta} 
R_{\gamma \varepsilon \epsilon}{}^{\theta} 
R_{\mu \theta \zeta \eta} + \frac{5}{4} 
H_{\alpha}{}^{\delta \epsilon} H^{\alpha \beta \gamma} 
H_{\beta}{}^{\varepsilon \mu} H_{\delta \varepsilon}{}^{\zeta} 
R_{\gamma}{}^{\eta}{}_{\epsilon}{}^{\theta} 
R_{\mu \theta \zeta \eta}\nn\\&& -  \frac{5}{32} 
H_{\alpha \beta}{}^{\delta} H^{\alpha \beta \gamma} 
H_{\gamma}{}^{\epsilon \varepsilon} H_{\delta \epsilon 
\varepsilon} R_{\mu \eta \zeta \theta} 
R^{\mu \zeta \eta \theta}\labell{R2H4}
\eeqa
In this case there are 17 terms   in the structure $R^2H^4$ in the scheme used in \cite{Garousi:2020gio}, whereas in the present scheme there are only 7 terms.

Having set all parameters in the structures $H^8$, $RH^6$ and $R^2H^4$ to  zero, except the couplings in \reef{H8}, \reef{RH6} and \reef{R2H4}, we now find the minimum number of couplings   with structure $R^3H^2$. To this end, we  again keep only the terms that if one set to zero any of them, the equation \reef{DLK} has no solution. In this case there are 22 terms. Setting all parameters in the structure $R^3H^2$  to zero, except those 22 parameters, and solving \reef{DLK}, one finds the following couplings in the structure $R^3H^2$: 
\beqa
L_3^{R^3H^2}&=&- \frac{15}{2} H^{\alpha \beta \gamma} H^{\delta \epsilon \
\varepsilon} R_{\alpha \delta}{}^{\mu \zeta} R_{\beta \mu 
\epsilon}{}^{\eta} R_{\gamma \eta \varepsilon \zeta} + 
\frac{17}{6} H^{\alpha \beta \gamma} H^{\delta \epsilon 
\varepsilon} R_{\alpha}{}^{\mu}{}_{\delta}{}^{\zeta} R_{\beta 
\zeta \epsilon}{}^{\eta} R_{\gamma \eta \varepsilon \mu}\nn\\&& -  
\frac{15}{2} H^{\alpha \beta \gamma} H^{\delta \epsilon 
\varepsilon} R_{\alpha}{}^{\mu}{}_{\beta}{}^{\zeta} 
R_{\gamma}{}^{\eta}{}_{\delta \mu} R_{\epsilon \zeta 
\varepsilon \eta} + \frac{53}{2} H^{\alpha \beta \gamma} 
H^{\delta \epsilon \varepsilon} 
R_{\alpha}{}^{\mu}{}_{\beta}{}^{\zeta} R_{\gamma \mu 
\delta}{}^{\eta} R_{\epsilon \zeta \varepsilon \eta}\nn\\&& - 14 
H_{\alpha}{}^{\delta \epsilon} H^{\alpha \beta \gamma} 
R_{\beta \delta}{}^{\varepsilon \mu} \
R_{\gamma}{}^{\zeta}{}_{\varepsilon}{}^{\eta} R_{\epsilon \zeta 
\mu \eta} + 30 H_{\alpha}{}^{\delta \epsilon} H^{\alpha \beta 
\gamma} R_{\beta}{}^{\varepsilon}{}_{\delta}{}^{\mu} 
R_{\gamma}{}^{\zeta}{}_{\varepsilon}{}^{\eta} R_{\epsilon \zeta 
\mu \eta}\nn\\&& + 2 H_{\alpha}{}^{\delta \epsilon} H^{\alpha \beta 
\gamma} R_{\beta}{}^{\varepsilon}{}_{\gamma}{}^{\mu} 
R_{\delta}{}^{\zeta}{}_{\varepsilon}{}^{\eta} R_{\epsilon \zeta 
\mu \eta} + 2 H_{\alpha}{}^{\delta \epsilon} H^{\alpha \beta 
\gamma} R_{\beta \delta}{}^{\varepsilon \mu} 
R_{\gamma}{}^{\zeta}{}_{\varepsilon}{}^{\eta} R_{\epsilon \eta 
\mu \zeta} \nn\\&&- 18 H_{\alpha}{}^{\delta \epsilon} H^{\alpha \beta 
\gamma} R_{\beta}{}^{\varepsilon}{}_{\delta}{}^{\mu} 
R_{\gamma}{}^{\zeta}{}_{\varepsilon}{}^{\eta} R_{\epsilon \eta 
\mu \zeta} + 14 H_{\alpha}{}^{\delta \epsilon} H^{\alpha \beta 
\gamma} R_{\beta}{}^{\varepsilon}{}_{\gamma}{}^{\mu} 
R_{\delta}{}^{\zeta}{}_{\varepsilon}{}^{\eta} R_{\epsilon \eta 
\mu \zeta}\nn\\&& + \frac{5}{2} H_{\alpha \beta}{}^{\delta} 
H^{\alpha \beta \gamma} R_{\gamma}{}^{\epsilon \varepsilon 
\mu} R_{\delta}{}^{\zeta}{}_{\varepsilon}{}^{\eta} R_{\epsilon 
\eta \mu \zeta} -  \frac{165}{2} H^{\alpha \beta \gamma} 
H^{\delta \epsilon \varepsilon} R_{\alpha \delta 
\beta}{}^{\mu} R_{\gamma}{}^{\zeta}{}_{\epsilon}{}^{\eta} 
R_{\varepsilon \zeta \mu \eta} \nn\\&&+ 5 H_{\alpha}{}^{\delta 
\epsilon} H^{\alpha \beta \gamma} R_{\beta}{}^{\varepsilon}{}_{
\delta}{}^{\mu} R_{\gamma}{}^{\zeta}{}_{\epsilon}{}^{\eta} 
R_{\varepsilon \zeta \mu \eta} + 17 H^{\alpha \beta \gamma} 
H^{\delta \epsilon \varepsilon} R_{\alpha \delta \beta 
\epsilon} R_{\gamma}{}^{\mu \zeta \eta} R_{\varepsilon \zeta 
\mu \eta} \nn\\&&-  \frac{5}{4} H_{\alpha \beta}{}^{\delta} 
H^{\alpha \beta \gamma} R_{\gamma}{}^{\epsilon \varepsilon 
\mu} R_{\delta \epsilon}{}^{\zeta \eta} R_{\varepsilon \zeta 
\mu \eta} - 9 H_{\alpha}{}^{\delta \epsilon} H^{\alpha \beta 
\gamma} R_{\beta}{}^{\varepsilon}{}_{\gamma}{}^{\mu} 
R_{\delta}{}^{\zeta}{}_{\epsilon}{}^{\eta} R_{\varepsilon \zeta 
\mu \eta}\nn\\&& + 8 H_{\alpha}{}^{\delta \epsilon} H^{\alpha \beta 
\gamma} R_{\beta \delta \gamma}{}^{\varepsilon} 
R_{\epsilon}{}^{\mu \zeta \eta} R_{\varepsilon \zeta \mu 
\eta} + \frac{5}{2} H_{\alpha \beta}{}^{\delta} H^{\alpha 
\beta \gamma} 
R_{\gamma}{}^{\epsilon}{}_{\delta}{}^{\varepsilon} 
R_{\epsilon}{}^{\mu \zeta \eta} R_{\varepsilon \zeta \mu 
\eta}\nn\\&& - 7 H^{\alpha \beta \gamma} H^{\delta \epsilon 
\varepsilon} R_{\alpha \beta}{}^{\mu \zeta} R_{\gamma \delta 
\epsilon}{}^{\eta} R_{\varepsilon \eta \mu \zeta} + 40 
H^{\alpha \beta \gamma} H^{\delta \epsilon \varepsilon} 
R_{\alpha \delta \beta}{}^{\mu} 
R_{\gamma}{}^{\zeta}{}_{\epsilon}{}^{\eta} R_{\varepsilon \eta 
\mu \zeta}\nn\\&& - 14 H_{\alpha}{}^{\delta \epsilon} H^{\alpha \beta 
\gamma} R_{\beta}{}^{\varepsilon}{}_{\delta}{}^{\mu} 
R_{\gamma}{}^{\zeta}{}_{\epsilon}{}^{\eta} R_{\varepsilon \eta 
\mu \zeta} + \frac{5}{4} H_{\alpha}{}^{\delta \epsilon} 
H^{\alpha \beta \gamma} R_{\beta \delta \gamma \epsilon} 
R_{\varepsilon \zeta \mu \eta} R^{\varepsilon \mu \zeta \eta}\labell{R3H2}
\eeqa
Note that there are 13 couplings  in the structure $R^3H^2$ in the scheme used in \cite{Garousi:2020gio}, whereas in the present scheme there are more couplings, \ie 22 couplings. Using the above couplings and the couplings in the structure $R^2(\nabla H)^2$ that we will find in a moment, one can calculate the  three-graviton-two-B-field S-matrix element. The result should be reproduced by the corresponding  terms in the low energy expansion, \ie   eight-momentum level, of the string theory S-matrix element of five NS-NS vertex operators. On the other hand, using the couplings in \cite{Garousi:2020gio}, one   should  also reproduce  exactly the same eight-momentum terms of the string theory  S-matrix element. It means  if one calculates, in field theory, the three-graviton-two-B-field S-matrix element  using  the above couplings and  the couplings in the structure $R^2(\nabla H)^2$, then the resulting  S-matrix element should be the same as  the  S-matrix element corresponding to the  couplings in \cite{Garousi:2020gio}. Such identity of S-matrix elements should exist for all couplings in the present scheme and the couplings in the scheme used in \cite{Garousi:2020gio}. We will check this identity for four-point functions which have only contact terms at eight-momentum level. All higher-point functions have both contact terms and massless poles.

Having set all parameters in the structures $H^8$, $RH^6$, $R^2H^4$ and $R^3H^2$ to  zero, except the couplings in \reef{H8}, \reef{RH6}, \reef{R2H4} and \reef{R3H2}, we now find the minimum number of couplings   with structure $(\nabla H)^2H^4$. To this end, we  again keep only the terms that if one set to zero any of them, the equation \reef{DLK} has no solution. In this case there are 77 terms. Setting all parameters in the structure $(\nabla H)^2H^4$ to zero , except those 77 parameters, and solving \reef{DLK}, one finds the following couplings in the structure $(\nabla H)^2H^4$: 
\beqa
L_3^{(\prt H)^2H^4}&\!\!\!\!\!\!=\!\!\!\!\!\!&\frac{5}{8} H_{\alpha}{}^{\delta \epsilon} H^{\alpha \beta \gamma} H_{\beta}{}^{\varepsilon \mu} H_{\delta \varepsilon}{}^{\zeta} \nabla_{\epsilon}H_{\gamma}{}^{\eta \theta} \nabla_{\zeta}H_{\mu \eta \theta} -  \frac{1}{12} H_{\alpha \beta}{}^{\delta} H^{\alpha \beta \gamma} H_{\epsilon}{}^{\zeta \eta} H^{\epsilon \varepsilon \mu} \nabla_{\zeta}H_{\gamma \varepsilon}{}^{\theta} \nabla_{\eta}H_{\delta \mu \theta}\nn\\&& -  \frac{7}{8} H_{\alpha}{}^{\delta \epsilon} H^{\alpha \beta \gamma} H_{\beta}{}^{\varepsilon \mu} H_{\delta}{}^{\zeta \eta} \nabla_{\varepsilon}H_{\gamma \zeta}{}^{\theta} \nabla_{\eta}H_{\epsilon \mu \theta} + \frac{3}{2} H_{\alpha}{}^{\delta \epsilon} H^{\alpha \beta \gamma} H_{\beta}{}^{\varepsilon \mu} H_{\delta}{}^{\zeta \eta} \nabla_{\zeta}H_{\gamma \varepsilon}{}^{\theta} \nabla_{\eta}H_{\epsilon \mu \theta}\nn\\&& + \frac{31}{96} H_{\alpha \beta}{}^{\delta} H^{\alpha \beta \gamma} H_{\epsilon}{}^{\zeta \eta} H^{\epsilon \varepsilon \mu} \nabla_{\delta}H_{\gamma \varepsilon}{}^{\theta} \nabla_{\eta}H_{\mu \zeta \theta} + \frac{7}{12} H_{\alpha}{}^{\delta \epsilon} H^{\alpha \beta \gamma} H_{\beta}{}^{\varepsilon \mu} H_{\delta}{}^{\zeta \eta} \nabla_{\epsilon}H_{\gamma \varepsilon}{}^{\theta} \nabla_{\eta}H_{\mu \zeta \theta}\nn\\&& -  \frac{23}{12} H_{\alpha}{}^{\delta \epsilon} H^{\alpha \beta \gamma} H_{\beta}{}^{\varepsilon \mu} H_{\delta}{}^{\zeta \eta} \nabla_{\varepsilon}H_{\gamma \epsilon}{}^{\theta} \nabla_{\eta}H_{\mu \zeta \theta} + \frac{1}{6} H_{\alpha}{}^{\delta \epsilon} H^{\alpha \beta \gamma} H_{\beta}{}^{\varepsilon \mu} H_{\gamma}{}^{\zeta \eta} \nabla_{\varepsilon}H_{\delta \epsilon}{}^{\theta} \nabla_{\eta}H_{\mu \zeta \theta}\nn\\&& -  \frac{3}{8} H_{\alpha}{}^{\delta \epsilon} H^{\alpha \beta \gamma} H_{\varepsilon}{}^{\eta \theta} H^{\varepsilon \mu \zeta} \nabla_{\eta}H_{\beta \delta \mu} \nabla_{\theta}H_{\gamma \epsilon \zeta} + \frac{43}{96} H_{\alpha}{}^{\delta \epsilon} H^{\alpha \beta \gamma} H_{\varepsilon}{}^{\eta \theta} H^{\varepsilon \mu \zeta} \nabla_{\zeta}H_{\beta \delta \mu} \nabla_{\theta}H_{\gamma \epsilon \eta}\nn\\&& + \frac{47}{192} H_{\alpha}{}^{\delta \epsilon} H^{\alpha \beta \gamma} H_{\varepsilon}{}^{\eta \theta} H^{\varepsilon \mu \zeta} \nabla_{\eta}H_{\beta \gamma \mu} \nabla_{\theta}H_{\delta \epsilon \zeta} -  \frac{9}{64} H_{\alpha}{}^{\delta \epsilon} H^{\alpha \beta \gamma} H_{\varepsilon}{}^{\eta \theta} H^{\varepsilon \mu \zeta} \nabla_{\zeta}H_{\beta \gamma \mu} \nabla_{\theta}H_{\delta \epsilon \eta} \nn\\&&+ \frac{91}{96} H_{\alpha \beta}{}^{\delta} H^{\alpha \beta \gamma} H^{\epsilon \varepsilon \mu} H^{\zeta \eta \theta} \nabla_{\varepsilon}H_{\gamma \epsilon \zeta} \nabla_{\theta}H_{\delta \mu \eta} + \frac{3}{64} H_{\alpha \beta}{}^{\delta} H^{\alpha \beta \gamma} H^{\epsilon \varepsilon \mu} H^{\zeta \eta \theta} \nabla_{\zeta}H_{\gamma \epsilon \varepsilon} \nabla_{\theta}H_{\delta \mu \eta}\nn\\&& + \frac{61}{96} H_{\alpha}{}^{\delta \epsilon} H^{\alpha \beta \gamma} H_{\beta}{}^{\varepsilon \mu} H^{\zeta \eta \theta} \nabla_{\delta}H_{\gamma \zeta \eta} \nabla_{\theta}H_{\epsilon \varepsilon \mu} + \frac{173}{96} H_{\alpha}{}^{\delta \epsilon} H^{\alpha \beta \gamma} H_{\beta}{}^{\varepsilon \mu} H^{\zeta \eta \theta} \nabla_{\eta}H_{\gamma \delta \zeta} \nabla_{\theta}H_{\epsilon \varepsilon \mu}\nn\\&& + \frac{1}{6} H_{\alpha}{}^{\delta \epsilon} H^{\alpha \beta \gamma} H_{\varepsilon}{}^{\eta \theta} H^{\varepsilon \mu \zeta} \nabla_{\delta}H_{\beta \gamma \mu} \nabla_{\theta}H_{\epsilon \zeta \eta} -  \frac{1}{12} H_{\alpha}{}^{\delta \epsilon} H^{\alpha \beta \gamma} H_{\beta}{}^{\varepsilon \mu} H^{\zeta \eta \theta} \nabla_{\varepsilon}H_{\gamma \delta \zeta} \nabla_{\theta}H_{\epsilon \mu \eta}\nn\\&& -  \frac{91}{192} H_{\alpha \beta}{}^{\delta} H^{\alpha \beta \gamma} H^{\epsilon \varepsilon \mu} H^{\zeta \eta \theta} \nabla_{\delta}H_{\gamma \epsilon \zeta} \nabla_{\theta}H_{\varepsilon \mu \eta} + \frac{121}{48} H_{\alpha}{}^{\delta \epsilon} H^{\alpha \beta \gamma} H_{\beta}{}^{\varepsilon \mu} H^{\zeta \eta \theta} \nabla_{\epsilon}H_{\gamma \delta \zeta} \nabla_{\theta}H_{\varepsilon \mu \eta}\nn\\&& -  \frac{107}{96} H_{\alpha}{}^{\delta \epsilon} H^{\alpha \beta \gamma} H_{\beta}{}^{\varepsilon \mu} H^{\zeta \eta \theta} \nabla_{\zeta}H_{\gamma \delta \epsilon} \nabla_{\theta}H_{\varepsilon \mu \eta} -  \frac{35}{192} H_{\alpha \beta}{}^{\delta} H^{\alpha \beta \gamma} H_{\epsilon}{}^{\zeta \eta} H^{\epsilon \varepsilon \mu} \nabla_{\delta}H_{\gamma \varepsilon}{}^{\theta} \nabla_{\theta}H_{\mu \zeta \eta}\nn\\&& + \frac{31}{96} H_{\alpha}{}^{\delta \epsilon} H^{\alpha \beta \gamma} H_{\beta}{}^{\varepsilon \mu} H^{\zeta \eta \theta} \nabla_{\epsilon}H_{\gamma \delta \varepsilon} \nabla_{\theta}H_{\mu \zeta \eta} -  \frac{37}{48} H_{\alpha}{}^{\delta \epsilon} H^{\alpha \beta \gamma} H_{\beta}{}^{\varepsilon \mu} H_{\delta}{}^{\zeta \eta} \nabla_{\epsilon}H_{\gamma \varepsilon}{}^{\theta} \nabla_{\theta}H_{\mu \zeta \eta}\nn\\&& -  \frac{35}{192} H_{\alpha}{}^{\delta \epsilon} H^{\alpha \beta \gamma} H_{\beta}{}^{\varepsilon \mu} H^{\zeta \eta \theta} \nabla_{\varepsilon}H_{\gamma \delta \epsilon} \nabla_{\theta}H_{\mu \zeta \eta} + \frac{23}{16} H_{\alpha}{}^{\delta \epsilon} H^{\alpha \beta \gamma} H_{\beta}{}^{\varepsilon \mu} H_{\delta}{}^{\zeta \eta} \nabla_{\varepsilon}H_{\gamma \epsilon}{}^{\theta} \nabla_{\theta}H_{\mu \zeta \eta} \nn\\&&-  \frac{5}{24} H_{\alpha \beta}{}^{\delta} H^{\alpha \beta \gamma} H_{\gamma}{}^{\epsilon \varepsilon} H^{\mu \zeta \eta} \nabla_{\varepsilon}H_{\delta \epsilon}{}^{\theta} \nabla_{\theta}H_{\mu \zeta \eta} -  \frac{7}{6} H_{\alpha \beta}{}^{\delta} H^{\alpha \beta \gamma} H_{\gamma}{}^{\epsilon \varepsilon} H_{\epsilon}{}^{\mu \zeta} \nabla_{\varepsilon}H_{\delta}{}^{\eta \theta} \nabla_{\theta}H_{\mu \zeta \eta}\nn\\&& + 2 H_{\alpha}{}^{\delta \epsilon} H^{\alpha \beta \gamma} H_{\beta}{}^{\varepsilon \mu} H_{\delta}{}^{\zeta \eta} \nabla_{\eta}H_{\mu \zeta \theta} \nabla^{\theta}H_{\gamma \epsilon \varepsilon} -  \frac{19}{16} H_{\alpha}{}^{\delta \epsilon} H^{\alpha \beta \gamma} H_{\beta}{}^{\varepsilon \mu} H_{\delta}{}^{\zeta \eta} \nabla_{\theta}H_{\mu \zeta \eta} \nabla^{\theta}H_{\gamma \epsilon \varepsilon} \nn\\&& + \frac{3}{8} H_{\alpha}{}^{\delta \epsilon} H^{\alpha \beta \gamma} H_{\beta}{}^{\varepsilon \mu} H_{\delta \varepsilon}{}^{\zeta} \nabla_{\eta}H_{\mu \zeta \theta} \nabla^{\theta}H_{\gamma \epsilon}{}^{\eta}-  \frac{3}{8} H_{\alpha}{}^{\delta \epsilon} H^{\alpha \beta \gamma} H_{\beta}{}^{\varepsilon \mu} H_{\delta \varepsilon}{}^{\zeta} \nabla_{\theta}H_{\mu \zeta \eta} \nabla^{\theta}H_{\gamma \epsilon}{}^{\eta} \nn\\&& -  \frac{5}{8} H_{\alpha}{}^{\delta \epsilon} H^{\alpha \beta \gamma} H_{\beta \delta}{}^{\varepsilon} H^{\mu \zeta \eta} \nabla_{\eta}H_{\varepsilon \zeta \theta} \nabla^{\theta}H_{\gamma \epsilon \mu} + \frac{1}{8} H_{\alpha}{}^{\delta \epsilon} H^{\alpha \beta \gamma} H_{\beta \delta}{}^{\varepsilon} H^{\mu \zeta \eta} \nabla_{\theta}H_{\varepsilon \zeta \eta} \nabla^{\theta}H_{\gamma \epsilon \mu} \nn\\&& + \frac{7}{12} H_{\alpha \beta}{}^{\delta} H^{\alpha \beta \gamma} H_{\epsilon}{}^{\zeta \eta} H^{\epsilon \varepsilon \mu} \nabla_{\eta}H_{\delta \mu \theta} \nabla^{\theta}H_{\gamma \varepsilon \zeta} -  \frac{3}{2} H_{\alpha}{}^{\delta \epsilon} H^{\alpha \beta \gamma} H_{\beta}{}^{\varepsilon \mu} H_{\delta}{}^{\zeta \eta} \nabla_{\eta}H_{\epsilon \mu \theta} \nabla^{\theta}H_{\gamma \varepsilon \zeta} \nn\\&& -  \frac{3}{16} H_{\alpha \beta}{}^{\delta} H^{\alpha \beta \gamma} H_{\epsilon}{}^{\zeta \eta} H^{\epsilon \varepsilon \mu} \nabla_{\theta}H_{\delta \mu \eta} \nabla^{\theta}H_{\gamma \varepsilon \zeta} + \frac{131}{192} H_{\alpha \beta}{}^{\delta} H^{\alpha \beta \gamma} H_{\epsilon}{}^{\zeta \eta} H^{\epsilon \varepsilon \mu} \nabla_{\eta}H_{\delta \zeta \theta} \nabla^{\theta}H_{\gamma \varepsilon \mu} \nn\\&& + \frac{57}{16} H_{\alpha}{}^{\delta \epsilon} H^{\alpha \beta \gamma} H_{\beta}{}^{\varepsilon \mu} H_{\delta}{}^{\zeta \eta} \nabla_{\eta}H_{\epsilon \zeta \theta} \nabla^{\theta}H_{\gamma \varepsilon \mu} -  \frac{143}{192} H_{\alpha}{}^{\delta \epsilon} H^{\alpha \beta \gamma} H_{\beta}{}^{\varepsilon \mu} H_{\delta}{}^{\zeta \eta} \nabla_{\theta}H_{\epsilon \zeta \eta} \nabla^{\theta}H_{\gamma \varepsilon \mu}  \nn\\&&+ \frac{13}{24} H_{\alpha}{}^{\delta \epsilon} H^{\alpha \beta \gamma} H_{\beta}{}^{\varepsilon \mu} H_{\delta}{}^{\zeta \eta} \nabla_{\theta}H_{\epsilon \varepsilon \mu} \nabla^{\theta}H_{\gamma \zeta \eta} -  \frac{2}{3} H_{\alpha}{}^{\delta \epsilon} H^{\alpha \beta \gamma} H_{\beta \delta}{}^{\varepsilon} H^{\mu \zeta \eta} \nabla_{\varepsilon}H_{\epsilon \eta \theta} \nabla^{\theta}H_{\gamma \mu \zeta} \nn\\&& -  \frac{3}{128} H_{\alpha \beta}{}^{\delta} H^{\alpha \beta \gamma} H_{\epsilon \varepsilon}{}^{\zeta} H^{\epsilon \varepsilon \mu} \nabla_{\eta}H_{\delta \zeta \theta} \nabla^{\theta}H_{\gamma \mu}{}^{\eta} + \frac{3}{128} H_{\alpha \beta}{}^{\delta} H^{\alpha \beta \gamma} H_{\epsilon \varepsilon}{}^{\zeta} H^{\epsilon \varepsilon \mu} \nabla_{\theta}H_{\delta \zeta \eta} \nabla^{\theta}H_{\gamma \mu}{}^{\eta} \nn\\&& -  \frac{43}{96} H_{\alpha}{}^{\delta \epsilon} H^{\alpha \beta \gamma} H_{\beta}{}^{\varepsilon \mu} H_{\gamma}{}^{\zeta \eta} \nabla_{\eta}H_{\mu \zeta \theta} \nabla^{\theta}H_{\delta \epsilon \varepsilon} + \frac{13}{192} H_{\alpha}{}^{\delta \epsilon} H^{\alpha \beta \gamma} H_{\beta}{}^{\varepsilon \mu} H_{\gamma}{}^{\zeta \eta} \nabla_{\theta}H_{\mu \zeta \eta} \nabla^{\theta}H_{\delta \epsilon \varepsilon} \nn\\&& -  \frac{13}{16} H_{\alpha \beta}{}^{\delta} H^{\alpha \beta \gamma} H_{\gamma}{}^{\epsilon \varepsilon} H_{\epsilon}{}^{\mu \zeta} \nabla_{\eta}H_{\mu \zeta \theta} \nabla^{\theta}H_{\delta \varepsilon}{}^{\eta} + \frac{15}{16} H_{\alpha \beta}{}^{\delta} H^{\alpha \beta \gamma} H_{\gamma}{}^{\epsilon \varepsilon} H_{\epsilon}{}^{\mu \zeta} \nabla_{\theta}H_{\mu \zeta \eta} \nabla^{\theta}H_{\delta \varepsilon}{}^{\eta} \nn\\&& + \frac{11}{16} H_{\alpha \beta}{}^{\delta} H^{\alpha \beta \gamma} H_{\gamma}{}^{\epsilon \varepsilon} H_{\epsilon \varepsilon}{}^{\mu} \nabla_{\eta}H_{\mu \zeta \theta} \nabla^{\theta}H_{\delta}{}^{\zeta \eta} -  \frac{11}{32} H_{\alpha \beta}{}^{\delta} H^{\alpha \beta \gamma} H_{\gamma}{}^{\epsilon \varepsilon} H^{\mu \zeta \eta} \nabla_{\eta}H_{\epsilon \varepsilon \theta} \nabla^{\theta}H_{\delta \mu \zeta} \nn\\&& + \frac{31}{96} H_{\alpha \beta}{}^{\delta} H^{\alpha \beta \gamma} H_{\gamma}{}^{\epsilon \varepsilon} H^{\mu \zeta \eta} \nabla_{\theta}H_{\epsilon \varepsilon \eta} \nabla^{\theta}H_{\delta \mu \zeta} -  \frac{313}{96} H_{\alpha \beta}{}^{\delta} H^{\alpha \beta \gamma} H_{\gamma}{}^{\epsilon \varepsilon} H_{\epsilon}{}^{\mu \zeta} \nabla_{\zeta}H_{\varepsilon \eta \theta} \nabla^{\theta}H_{\delta \mu}{}^{\eta} \nn\\&& -  \frac{13}{12} H_{\alpha \beta}{}^{\delta} H^{\alpha \beta \gamma} H_{\gamma}{}^{\epsilon \varepsilon} H_{\epsilon}{}^{\mu \zeta} \nabla_{\eta}H_{\varepsilon \zeta \theta} \nabla^{\theta}H_{\delta \mu}{}^{\eta} + \frac{13}{12} H_{\alpha \beta}{}^{\delta} H^{\alpha \beta \gamma} H_{\gamma}{}^{\epsilon \varepsilon} H_{\epsilon}{}^{\mu \zeta} \nabla_{\theta}H_{\varepsilon \zeta \eta} \nabla^{\theta}H_{\delta \mu}{}^{\eta} \nn\\&& -  \frac{13}{48} H_{\alpha}{}^{\delta \epsilon} H^{\alpha \beta \gamma} H_{\beta \delta}{}^{\varepsilon} H_{\gamma}{}^{\mu \zeta} \nabla_{\eta}H_{\mu \zeta \theta} \nabla^{\theta}H_{\epsilon \varepsilon}{}^{\eta} + \frac{7}{48} H_{\alpha}{}^{\delta \epsilon} H^{\alpha \beta \gamma} H_{\beta \delta}{}^{\varepsilon} H_{\gamma}{}^{\mu \zeta} \nabla_{\theta}H_{\mu \zeta \eta} \nabla^{\theta}H_{\epsilon \varepsilon}{}^{\eta} \nn\\&& -  \frac{5}{32} H_{\alpha \beta}{}^{\delta} H^{\alpha \beta \gamma} H_{\gamma}{}^{\epsilon \varepsilon} H_{\delta}{}^{\mu \zeta} \nabla_{\theta}H_{\mu \zeta \eta} \nabla^{\theta}H_{\epsilon \varepsilon}{}^{\eta} + \frac{17}{96} H_{\alpha \beta}{}^{\delta} H^{\alpha \beta \gamma} H_{\gamma}{}^{\epsilon \varepsilon} H^{\mu \zeta \eta} \nabla_{\delta}H_{\zeta \eta \theta} \nabla^{\theta}H_{\epsilon \varepsilon \mu}  \nn\\&&+ \frac{5}{4} H_{\alpha \beta}{}^{\delta} H^{\alpha \beta \gamma} H_{\gamma}{}^{\epsilon \varepsilon} H^{\mu \zeta \eta} \nabla_{\delta}H_{\varepsilon \eta \theta} \nabla^{\theta}H_{\epsilon \mu \zeta} + \frac{27}{32} H_{\alpha \beta}{}^{\delta} H^{\alpha \beta \gamma} H_{\gamma}{}^{\epsilon \varepsilon} H_{\delta}{}^{\mu \zeta} \nabla_{\eta}H_{\varepsilon \zeta \theta} \nabla^{\theta}H_{\epsilon \mu}{}^{\eta} \nn\\&& + \frac{119}{192} H_{\alpha \beta}{}^{\delta} H^{\alpha \beta \gamma} H_{\gamma}{}^{\epsilon \varepsilon} H_{\delta}{}^{\mu \zeta} \nabla_{\theta}H_{\varepsilon \zeta \eta} \nabla^{\theta}H_{\epsilon \mu}{}^{\eta} -  \frac{3}{8} H_{\alpha}{}^{\delta \epsilon} H^{\alpha \beta \gamma} H_{\beta \delta}{}^{\varepsilon} H_{\gamma \epsilon}{}^{\mu} \nabla_{\eta}H_{\mu \zeta \theta} \nabla^{\theta}H_{\varepsilon}{}^{\zeta \eta} \nn\\&& + \frac{383}{96} H_{\alpha \beta}{}^{\delta} H^{\alpha \beta \gamma} H_{\gamma}{}^{\epsilon \varepsilon} H_{\delta \epsilon}{}^{\mu} \nabla_{\eta}H_{\mu \zeta \theta} \nabla^{\theta}H_{\varepsilon}{}^{\zeta \eta} + \frac{3}{16} H_{\alpha}{}^{\delta \epsilon} H^{\alpha \beta \gamma} H_{\beta \delta}{}^{\varepsilon} H_{\gamma \epsilon}{}^{\mu} \nabla_{\theta}H_{\mu \zeta \eta} \nabla^{\theta}H_{\varepsilon}{}^{\zeta \eta} \nn\\&& -  \frac{9}{16} H_{\alpha \beta}{}^{\delta} H^{\alpha \beta \gamma} H_{\gamma}{}^{\epsilon \varepsilon} H_{\delta \epsilon}{}^{\mu} \nabla_{\theta}H_{\mu \zeta \eta} \nabla^{\theta}H_{\varepsilon}{}^{\zeta \eta} + \frac{5}{12} H_{\alpha \beta}{}^{\delta} H^{\alpha \beta \gamma} H_{\gamma}{}^{\epsilon \varepsilon} H_{\epsilon}{}^{\mu \zeta} \nabla_{\delta}H_{\zeta \eta \theta} \nabla^{\theta}H_{\varepsilon \mu}{}^{\eta} \nn\\&& -  \frac{11}{48} H_{\alpha}{}^{\delta \epsilon} H^{\alpha \beta \gamma} H_{\varepsilon}{}^{\eta \theta} H^{\varepsilon \mu \zeta} \nabla_{\zeta}H_{\epsilon \eta \theta} \nabla_{\mu}H_{\beta \gamma \delta} + \frac{1}{8} H_{\alpha}{}^{\delta \epsilon} H^{\alpha \beta \gamma} H_{\beta \delta}{}^{\varepsilon} H^{\mu \zeta \eta} \nabla_{\eta}H_{\varepsilon \zeta \theta} \nabla_{\mu}H_{\gamma \epsilon}{}^{\theta} \nn\\&& -  \frac{61}{96} H_{\alpha \beta}{}^{\delta} H^{\alpha \beta \gamma} H_{\epsilon}{}^{\zeta \eta} H^{\epsilon \varepsilon \mu} \nabla_{\eta}H_{\delta \zeta \theta} \nabla_{\mu}H_{\gamma \varepsilon}{}^{\theta} -  \frac{199}{48} H_{\alpha}{}^{\delta \epsilon} H^{\alpha \beta \gamma} H_{\beta}{}^{\varepsilon \mu} H_{\delta}{}^{\zeta \eta} \nabla_{\eta}H_{\epsilon \zeta \theta} \nabla_{\mu}H_{\gamma \varepsilon}{}^{\theta} \nn\\&& -  \frac{91}{192} H_{\alpha}{}^{\delta \epsilon} H^{\alpha \beta \gamma} H_{\beta}{}^{\varepsilon \mu} H^{\zeta \eta \theta} \nabla_{\theta}H_{\gamma \zeta \eta} \nabla_{\mu}H_{\delta \epsilon \varepsilon} -  \frac{5}{16} H_{\alpha \beta}{}^{\delta} H^{\alpha \beta \gamma} H_{\epsilon}{}^{\zeta \eta} H^{\epsilon \varepsilon \mu} \nabla_{\zeta}H_{\gamma \varepsilon}{}^{\theta} \nabla_{\mu}H_{\delta \eta \theta}  \nn\\&&-  \frac{53}{24} H_{\alpha}{}^{\delta \epsilon} H^{\alpha \beta \gamma} H_{\beta}{}^{\varepsilon \mu} H^{\zeta \eta \theta} \nabla_{\eta}H_{\gamma \delta \zeta} \nabla_{\mu}H_{\epsilon \varepsilon \theta} -  \frac{1}{16} H_{\alpha}{}^{\delta \epsilon} H^{\alpha \beta \gamma} H_{\beta}{}^{\varepsilon \mu} H_{\delta}{}^{\zeta \eta} \nabla_{\eta}H_{\gamma \zeta}{}^{\theta} \nabla_{\mu}H_{\epsilon \varepsilon \theta} \nn\\&& -  \frac{77}{96} H_{\alpha}{}^{\delta \epsilon} H^{\alpha \beta \gamma} H_{\beta}{}^{\varepsilon \mu} H_{\delta}{}^{\zeta \eta} \nabla^{\theta}H_{\gamma \zeta \eta} \nabla_{\mu}H_{\epsilon \varepsilon \theta} -  \frac{5}{16} H_{\alpha}{}^{\delta \epsilon} H^{\alpha \beta \gamma} H_{\beta}{}^{\varepsilon \mu} H_{\delta \varepsilon}{}^{\zeta} \nabla_{\zeta}H_{\gamma}{}^{\eta \theta} \nabla_{\mu}H_{\epsilon \eta \theta} \nn\\&& -  \frac{1}{2} H_{\alpha}{}^{\delta \epsilon} H^{\alpha \beta \gamma} H_{\beta}{}^{\varepsilon \mu} H_{\gamma}{}^{\zeta \eta} \nabla_{\zeta}H_{\delta \varepsilon}{}^{\theta} \nabla_{\mu}H_{\epsilon \eta \theta}\labell{PH2H4}
\eeqa
The corresponding Lagrangian in the scheme used in \cite{Garousi:2020gio} has  78 couplings.

Having set all parameters in the structures $H^8$, $RH^6$, $R^2H^4$,  $R^3H^2$, and $(\nabla H)^2H^4$ to  zero, except the couplings in \reef{H8}, \reef{RH6}, \reef{R2H4}, \reef{R3H2} and \reef{PH2H4}, we now find the minimum number of couplings   with structure $R(\nabla H)^2H^2$. To this end, we  again keep only the terms that if one set to zero any of them, the equation \reef{DLK} has no solution. In this case there are 106 terms. Setting all parameters in the structure $R(\nabla H)^2H^2$ to zero, except those 106 parameters,  and solving \reef{DLK}, one finds the following couplings in the structure $R(\nabla H)^2H^2$: 
\beqa
L_3^{R(\prt H)^2 H^2}&=&\frac{457}{48} H_{\alpha}{}^{\delta \epsilon} H^{\alpha \beta 
\gamma} R_{\varepsilon \zeta \mu \eta} 
\nabla_{\delta}H_{\beta}{}^{\varepsilon \mu} 
\nabla_{\epsilon}H_{\gamma}{}^{\zeta \eta} -  \frac{503}{24} 
H_{\alpha}{}^{\delta \epsilon} H^{\alpha \beta \gamma} 
R_{\gamma \zeta \mu \eta} 
\nabla_{\delta}H_{\beta}{}^{\varepsilon \mu} 
\nabla_{\varepsilon}H_{\epsilon}{}^{\zeta \eta}\nn\\&& + 
\frac{121}{24} H^{\alpha \beta \gamma} H^{\delta \epsilon 
\varepsilon} R_{\beta \varepsilon \gamma \eta} 
\nabla_{\delta}H_{\alpha}{}^{\mu \zeta} 
\nabla_{\zeta}H_{\epsilon \mu}{}^{\eta} + \frac{121}{48} 
H^{\alpha \beta \gamma} H^{\delta \epsilon \varepsilon} 
R_{\epsilon \mu \varepsilon \eta} \nabla_{\delta}H_{\gamma 
\zeta}{}^{\eta} \nabla^{\zeta}H_{\alpha \beta}{}^{\mu} \nn\\&&-  
\frac{25}{24} H^{\alpha \beta \gamma} H^{\delta \epsilon 
\varepsilon} R_{\epsilon \zeta \varepsilon \eta} 
\nabla_{\delta}H_{\gamma \mu}{}^{\eta} 
\nabla^{\zeta}H_{\alpha \beta}{}^{\mu} + \frac{169}{48} 
H^{\alpha \beta \gamma} H^{\delta \epsilon \varepsilon} 
R_{\epsilon \mu \varepsilon \eta} \nabla_{\zeta}H_{\gamma 
\delta}{}^{\eta} \nabla^{\zeta}H_{\alpha \beta}{}^{\mu}\nn\\&& -  
\frac{73}{48} H^{\alpha \beta \gamma} H^{\delta \epsilon 
\varepsilon} R_{\gamma \eta \varepsilon \mu} \nabla_{\zeta}H_{
\delta \epsilon}{}^{\eta} \nabla^{\zeta}H_{\alpha 
\beta}{}^{\mu} + \frac{67}{48} H^{\alpha \beta \gamma} 
H^{\delta \epsilon \varepsilon} R_{\gamma \mu \varepsilon 
\eta} \nabla_{\zeta}H_{\delta \epsilon}{}^{\eta} 
\nabla^{\zeta}H_{\alpha \beta}{}^{\mu}\nn\\&& + \frac{127}{8} 
H^{\alpha \beta \gamma} H^{\delta \epsilon \varepsilon} 
R_{\gamma \eta \epsilon \varepsilon} \nabla_{\zeta}H_{\delta 
\mu}{}^{\eta} \nabla^{\zeta}H_{\alpha \beta}{}^{\mu} + 
\frac{143}{12} H^{\alpha \beta \gamma} H^{\delta \epsilon 
\varepsilon} R_{\epsilon \mu \varepsilon \eta} 
\nabla_{\gamma}H_{\beta \zeta}{}^{\eta} 
\nabla^{\zeta}H_{\alpha \delta}{}^{\mu} \nn\\&&+ \frac{11}{12} 
H^{\alpha \beta \gamma} H^{\delta \epsilon \varepsilon} 
R_{\gamma \eta \varepsilon \mu} \nabla_{\epsilon}H_{\beta 
\zeta}{}^{\eta} \nabla^{\zeta}H_{\alpha \delta}{}^{\mu} + 
\frac{179}{6} H^{\alpha \beta \gamma} H^{\delta \epsilon 
\varepsilon} R_{\gamma \eta \varepsilon \zeta} 
\nabla_{\epsilon}H_{\beta \mu}{}^{\eta} 
\nabla^{\zeta}H_{\alpha \delta}{}^{\mu}\nn\\&& -  \frac{181}{24} 
H^{\alpha \beta \gamma} H^{\delta \epsilon \varepsilon} 
R_{\gamma \mu \varepsilon \eta} \nabla_{\zeta}H_{\beta 
\epsilon}{}^{\eta} \nabla^{\zeta}H_{\alpha \delta}{}^{\mu} -  
\frac{49}{6} H^{\alpha \beta \gamma} H^{\delta \epsilon 
\varepsilon} R_{\gamma \eta \epsilon \varepsilon} 
\nabla_{\zeta}H_{\beta \mu}{}^{\eta} \nabla^{\zeta}H_{\alpha 
\delta}{}^{\mu}\nn\\&& -  \frac{143}{8} H_{\alpha}{}^{\delta 
\epsilon} H^{\alpha \beta \gamma} R_{\delta \mu \epsilon 
\eta} \nabla_{\varepsilon}H_{\gamma \zeta}{}^{\eta} 
\nabla^{\zeta}H_{\beta}{}^{\varepsilon \mu} -  \frac{193}{24} 
H_{\alpha}{}^{\delta \epsilon} H^{\alpha \beta \gamma} 
R_{\gamma \eta \epsilon \mu} \nabla_{\varepsilon}H_{\delta 
\zeta}{}^{\eta} \nabla^{\zeta}H_{\beta}{}^{\varepsilon \mu} \nn\\&&+ 
\frac{73}{8} H_{\alpha}{}^{\delta \epsilon} H^{\alpha \beta 
\gamma} R_{\gamma \mu \epsilon \eta} 
\nabla_{\varepsilon}H_{\delta \zeta}{}^{\eta} 
\nabla^{\zeta}H_{\beta}{}^{\varepsilon \mu} + \frac{139}{8} 
H_{\alpha}{}^{\delta \epsilon} H^{\alpha \beta \gamma} 
R_{\delta \mu \epsilon \eta} \nabla_{\zeta}H_{\gamma 
\varepsilon}{}^{\eta} \nabla^{\zeta}H_{\beta}{}^{\varepsilon 
\mu}\nn\\&& -  \frac{61}{24} H_{\alpha}{}^{\delta \epsilon} H^{\alpha 
\beta \gamma} R_{\gamma \eta \epsilon \mu} 
\nabla_{\zeta}H_{\delta \varepsilon}{}^{\eta} 
\nabla^{\zeta}H_{\beta}{}^{\varepsilon \mu} -  \frac{23}{8} 
H_{\alpha}{}^{\delta \epsilon} H^{\alpha \beta \gamma} 
R_{\gamma \mu \epsilon \eta} \nabla_{\zeta}H_{\delta 
\varepsilon}{}^{\eta} \nabla^{\zeta}H_{\beta}{}^{\varepsilon 
\mu} \nn\\&&+ \frac{287}{48} H_{\alpha}{}^{\delta \epsilon} H^{\alpha 
\beta \gamma} R_{\gamma \eta \delta \epsilon} 
\nabla_{\zeta}H_{\varepsilon \mu}{}^{\eta} 
\nabla^{\zeta}H_{\beta}{}^{\varepsilon \mu} + \frac{5}{4} 
H_{\alpha \beta}{}^{\delta} H^{\alpha \beta \gamma} R_{\gamma 
\mu \delta \eta} \nabla_{\zeta}H_{\epsilon 
\varepsilon}{}^{\eta} \nabla^{\zeta}H^{\epsilon \varepsilon 
\mu}\nn\\&& -  \frac{121}{48} H^{\alpha \beta \gamma} H^{\delta 
\epsilon \varepsilon} R_{\beta \epsilon \gamma \varepsilon} 
\nabla_{\zeta}H_{\delta \mu \eta} \nabla^{\eta}H_{\alpha}{}^{
\mu \zeta} -  \frac{143}{32} H^{\alpha \beta \gamma} 
H^{\delta \epsilon \varepsilon} R_{\beta \epsilon \gamma 
\varepsilon} \nabla_{\eta}H_{\delta \mu \zeta} 
\nabla^{\eta}H_{\alpha}{}^{\mu \zeta} \nn\\&&-  \frac{131}{8} 
H^{\alpha \beta \gamma} H^{\delta \epsilon \varepsilon} 
R_{\gamma \mu \varepsilon \eta} \nabla^{\zeta}H_{\alpha 
\delta}{}^{\mu} \nabla^{\eta}H_{\beta \epsilon \zeta} + 
\frac{275}{24} H^{\alpha \beta \gamma} H^{\delta \epsilon 
\varepsilon} R_{\gamma \zeta \varepsilon \eta} 
\nabla^{\zeta}H_{\alpha \delta}{}^{\mu} \nabla^{\eta}H_{\beta 
\epsilon \mu}\nn\\&& + \frac{20}{3} H^{\alpha \beta \gamma} 
H^{\delta \epsilon \varepsilon} R_{\gamma \eta \epsilon 
\varepsilon} \nabla^{\zeta}H_{\alpha \delta}{}^{\mu} 
\nabla^{\eta}H_{\beta \mu \zeta} -  \frac{35}{48} H^{\alpha 
\beta \gamma} H^{\delta \epsilon \varepsilon} R_{\epsilon \mu 
\varepsilon \eta} \nabla^{\zeta}H_{\alpha \beta}{}^{\mu} 
\nabla^{\eta}H_{\gamma \delta \zeta}\nn\\&& -  \frac{67}{24} 
H^{\alpha \beta \gamma} H^{\delta \epsilon \varepsilon} 
R_{\epsilon \zeta \varepsilon \eta} \nabla^{\zeta}H_{\alpha 
\beta}{}^{\mu} \nabla^{\eta}H_{\gamma \delta \mu} + 
\frac{145}{12} H^{\alpha \beta \gamma} H^{\delta \epsilon 
\varepsilon} R_{\varepsilon \zeta \mu \eta} \nabla_{\delta}H_{
\alpha \beta}{}^{\mu} \nabla^{\eta}H_{\gamma 
\epsilon}{}^{\zeta} \nn\\&&-  \frac{73}{12} H^{\alpha \beta \gamma} 
H^{\delta \epsilon \varepsilon} R_{\varepsilon \eta \mu \zeta} 
\nabla_{\delta}H_{\alpha \beta}{}^{\mu} 
\nabla^{\eta}H_{\gamma \epsilon}{}^{\zeta} -  \frac{599}{24} 
H_{\alpha}{}^{\delta \epsilon} H^{\alpha \beta \gamma} 
R_{\delta \mu \epsilon \eta} 
\nabla^{\zeta}H_{\beta}{}^{\varepsilon \mu} 
\nabla^{\eta}H_{\gamma \varepsilon \zeta}\nn\\&& + \frac{61}{24} 
H_{\alpha}{}^{\delta \epsilon} H^{\alpha \beta \gamma} 
R_{\delta \zeta \epsilon \eta} 
\nabla^{\zeta}H_{\beta}{}^{\varepsilon \mu} 
\nabla^{\eta}H_{\gamma \varepsilon \mu} -  \frac{35}{4} 
H^{\alpha \beta \gamma} H^{\delta \epsilon \varepsilon} 
R_{\epsilon \zeta \varepsilon \eta} \nabla_{\delta}H_{\alpha 
\beta}{}^{\mu} \nabla^{\eta}H_{\gamma \mu}{}^{\zeta}\nn\\&& + 
\frac{73}{96} H^{\alpha \beta \gamma} H^{\delta \epsilon 
\varepsilon} R_{\gamma \eta \varepsilon \mu} \nabla^{\zeta}H_{
\alpha \beta}{}^{\mu} \nabla^{\eta}H_{\delta \epsilon \zeta} 
-  \frac{409}{96} H^{\alpha \beta \gamma} H^{\delta \epsilon 
\varepsilon} R_{\gamma \mu \varepsilon \eta} \nabla^{\zeta}H_{
\alpha \beta}{}^{\mu} \nabla^{\eta}H_{\delta \epsilon \zeta} 
\nn\\&&+ \frac{215}{96} H^{\alpha \beta \gamma} H^{\delta \epsilon 
\varepsilon} R_{\gamma \zeta \varepsilon \eta} 
\nabla^{\zeta}H_{\alpha \beta}{}^{\mu} \nabla^{\eta}H_{\delta 
\epsilon \mu} + \frac{13}{96} H^{\alpha \beta \gamma} 
H^{\delta \epsilon \varepsilon} R_{\gamma \eta \varepsilon 
\zeta} \nabla^{\zeta}H_{\alpha \beta}{}^{\mu} 
\nabla^{\eta}H_{\delta \epsilon \mu}\nn\\&& -  \frac{23}{2} 
H_{\alpha}{}^{\delta \epsilon} H^{\alpha \beta \gamma} 
R_{\gamma \eta \epsilon \mu} 
\nabla^{\zeta}H_{\beta}{}^{\varepsilon \mu} 
\nabla^{\eta}H_{\delta \varepsilon \zeta} -  \frac{103}{12} 
H_{\alpha}{}^{\delta \epsilon} H^{\alpha \beta \gamma} 
R_{\gamma \mu \epsilon \eta} 
\nabla^{\zeta}H_{\beta}{}^{\varepsilon \mu} 
\nabla^{\eta}H_{\delta \varepsilon \zeta}\nn\\&& + \frac{1}{4} 
H_{\alpha}{}^{\delta \epsilon} H^{\alpha \beta \gamma} 
R_{\gamma \zeta \epsilon \eta} 
\nabla^{\zeta}H_{\beta}{}^{\varepsilon \mu} 
\nabla^{\eta}H_{\delta \varepsilon \mu} + \frac{31}{24} 
H_{\alpha}{}^{\delta \epsilon} H^{\alpha \beta \gamma} 
R_{\gamma \eta \epsilon \zeta} 
\nabla^{\zeta}H_{\beta}{}^{\varepsilon \mu} 
\nabla^{\eta}H_{\delta \varepsilon \mu}\nn\\&& -  \frac{91}{12} 
H^{\alpha \beta \gamma} H^{\delta \epsilon \varepsilon} 
R_{\gamma \eta \epsilon \varepsilon} \nabla^{\zeta}H_{\alpha 
\beta}{}^{\mu} \nabla^{\eta}H_{\delta \mu \zeta} -  
\frac{5}{4} H_{\alpha \beta}{}^{\delta} H^{\alpha \beta 
\gamma} R_{\gamma \mu \delta \eta} \nabla^{\zeta}H^{\epsilon 
\varepsilon \mu} \nabla^{\eta}H_{\epsilon \varepsilon \zeta}\nn\\&& - 
 \frac{1}{8} H^{\alpha \beta \gamma} H^{\delta \epsilon 
\varepsilon} R_{\gamma \zeta \mu \eta} 
\nabla_{\delta}H_{\alpha \beta}{}^{\mu} 
\nabla^{\eta}H_{\epsilon \varepsilon}{}^{\zeta} -  
\frac{179}{96} H^{\alpha \beta \gamma} H^{\delta \epsilon 
\varepsilon} R_{\gamma \eta \mu \zeta} 
\nabla_{\delta}H_{\alpha \beta}{}^{\mu} 
\nabla^{\eta}H_{\epsilon \varepsilon}{}^{\zeta}\nn\\&& + 
\frac{83}{12} H_{\alpha}{}^{\delta \epsilon} H^{\alpha \beta 
\gamma} R_{\gamma \zeta \mu \eta} 
\nabla_{\delta}H_{\beta}{}^{\varepsilon \mu} 
\nabla^{\eta}H_{\epsilon \varepsilon}{}^{\zeta} + 
\frac{157}{24} H_{\alpha}{}^{\delta \epsilon} H^{\alpha \beta 
\gamma} R_{\gamma \eta \mu \zeta} 
\nabla_{\delta}H_{\beta}{}^{\varepsilon \mu} 
\nabla^{\eta}H_{\epsilon \varepsilon}{}^{\zeta}\nn\\&& + 
\frac{317}{48} H^{\alpha \beta \gamma} H^{\delta \epsilon 
\varepsilon} R_{\beta \varepsilon \gamma \eta} 
\nabla_{\delta}H_{\alpha}{}^{\mu \zeta} 
\nabla^{\eta}H_{\epsilon \mu \zeta} -  \frac{63}{8} H^{\alpha 
\beta \gamma} H^{\delta \epsilon \varepsilon} R_{\gamma \zeta 
\varepsilon \eta} \nabla_{\delta}H_{\alpha \beta}{}^{\mu} 
\nabla^{\eta}H_{\epsilon \mu}{}^{\zeta}\nn\\&& + \frac{79}{12} 
H^{\alpha \beta \gamma} H^{\delta \epsilon \varepsilon} 
R_{\gamma \eta \varepsilon \zeta} \nabla_{\delta}H_{\alpha 
\beta}{}^{\mu} \nabla^{\eta}H_{\epsilon \mu}{}^{\zeta} + 
\frac{155}{288} H^{\alpha \beta \gamma} H^{\delta \epsilon 
\varepsilon} R_{\varepsilon \eta \mu \zeta} \nabla_{\delta}H_{
\alpha \beta \gamma} \nabla^{\eta}H_{\epsilon}{}^{\mu \zeta} 
\nn\\&&-  \frac{311}{96} H_{\alpha}{}^{\delta \epsilon} H^{\alpha 
\beta \gamma} R_{\varepsilon \eta \mu \zeta} 
\nabla_{\delta}H_{\beta \gamma}{}^{\varepsilon} 
\nabla^{\eta}H_{\epsilon}{}^{\mu \zeta} + \frac{35}{96} 
H_{\alpha \beta}{}^{\delta} H^{\alpha \beta \gamma} 
R_{\varepsilon \eta \mu \zeta} 
\nabla_{\delta}H_{\gamma}{}^{\epsilon \varepsilon} 
\nabla^{\eta}H_{\epsilon}{}^{\mu \zeta}\nn\\&& + \frac{5}{8} 
H_{\alpha}{}^{\delta \epsilon} H^{\alpha \beta \gamma} 
R_{\varepsilon \eta \mu \zeta} \nabla^{\varepsilon}H_{\beta 
\gamma \delta} \nabla^{\eta}H_{\epsilon}{}^{\mu \zeta} -  
\frac{15}{2} H_{\alpha}{}^{\delta \epsilon} H^{\alpha \beta 
\gamma} R_{\gamma \eta \delta \epsilon} 
\nabla^{\zeta}H_{\beta}{}^{\varepsilon \mu} 
\nabla^{\eta}H_{\varepsilon \mu \zeta}\nn\\&& -  \frac{73}{24} 
H_{\alpha}{}^{\delta \epsilon} H^{\alpha \beta \gamma} 
R_{\gamma \eta \epsilon \zeta} 
\nabla_{\delta}H_{\beta}{}^{\varepsilon \mu} 
\nabla^{\eta}H_{\varepsilon \mu}{}^{\zeta} + \frac{287}{96} 
H_{\alpha}{}^{\delta \epsilon} H^{\alpha \beta \gamma} 
R_{\epsilon \eta \mu \zeta} \nabla_{\delta}H_{\beta 
\gamma}{}^{\varepsilon} \nabla^{\eta}H_{\varepsilon}{}^{\mu 
\zeta}\nn\\&& -  \frac{25}{48} H^{\alpha \beta \gamma} H^{\delta 
\epsilon \varepsilon} R_{\gamma \eta \mu \zeta} 
\nabla_{\epsilon}H_{\alpha \beta \delta} 
\nabla^{\eta}H_{\varepsilon}{}^{\mu \zeta} + \frac{1}{4} 
H_{\alpha}{}^{\delta \epsilon} H^{\alpha \beta \gamma} 
R_{\epsilon \eta \mu \zeta} \nabla^{\varepsilon}H_{\beta 
\gamma \delta} \nabla^{\eta}H_{\varepsilon}{}^{\mu \zeta}\nn\\&& + 
\frac{247}{48} H^{\alpha \beta \gamma} H^{\delta \epsilon 
\varepsilon} R_{\gamma \eta \epsilon \varepsilon} 
\nabla^{\zeta}H_{\alpha \delta}{}^{\mu} \nabla_{\mu}H_{\beta 
\zeta}{}^{\eta} -  \frac{45}{4} H^{\alpha \beta \gamma} 
H^{\delta \epsilon \varepsilon} R_{\epsilon \zeta \varepsilon 
\eta} \nabla_{\delta}H_{\alpha \beta}{}^{\mu} \nabla_{\mu}H_{
\gamma}{}^{\zeta \eta}\nn\\&& -  \frac{461}{32} H^{\alpha \beta 
\gamma} H^{\delta \epsilon \varepsilon} R_{\gamma \eta 
\epsilon \varepsilon} \nabla^{\zeta}H_{\alpha \beta}{}^{\mu} 
\nabla_{\mu}H_{\delta \zeta}{}^{\eta} + \frac{43}{48} 
H^{\alpha \beta \gamma} H^{\delta \epsilon \varepsilon} 
R_{\gamma \zeta \varepsilon \eta} \nabla_{\delta}H_{\alpha 
\beta}{}^{\mu} \nabla_{\mu}H_{\epsilon}{}^{\zeta \eta} \nn\\&&-  
\frac{5}{24} H^{\alpha \beta \gamma} H^{\delta \epsilon 
\varepsilon} R_{\varepsilon \zeta \mu \eta} 
\nabla^{\eta}H_{\delta \epsilon}{}^{\zeta} 
\nabla^{\mu}H_{\alpha \beta \gamma} + \frac{5}{6} H^{\alpha 
\beta \gamma} H^{\delta \epsilon \varepsilon} R_{\epsilon 
\zeta \varepsilon \eta} \nabla^{\eta}H_{\delta \mu}{}^{\zeta} 
\nabla^{\mu}H_{\alpha \beta \gamma} \nn\\&&+ \frac{5}{6} H^{\alpha 
\beta \gamma} H^{\delta \epsilon \varepsilon} R_{\epsilon 
\zeta \varepsilon \eta} \nabla_{\mu}H_{\delta}{}^{\zeta \eta} 
\nabla^{\mu}H_{\alpha \beta \gamma} -  \frac{167}{12} 
H^{\alpha \beta \gamma} H^{\delta \epsilon \varepsilon} 
R_{\varepsilon \zeta \mu \eta} \nabla_{\epsilon}H_{\gamma}{}^{
\zeta \eta} \nabla^{\mu}H_{\alpha \beta \delta}\nn\\&& -  
\frac{257}{16} H^{\alpha \beta \gamma} H^{\delta \epsilon 
\varepsilon} R_{\varepsilon \zeta \mu \eta} 
\nabla^{\eta}H_{\gamma \epsilon}{}^{\zeta} 
\nabla^{\mu}H_{\alpha \beta \delta} + \frac{16}{3} H^{\alpha 
\beta \gamma} H^{\delta \epsilon \varepsilon} R_{\varepsilon 
\eta \mu \zeta} \nabla^{\eta}H_{\gamma \epsilon}{}^{\zeta} 
\nabla^{\mu}H_{\alpha \beta \delta}\nn\\&& + \frac{50}{3} H^{\alpha 
\beta \gamma} H^{\delta \epsilon \varepsilon} R_{\epsilon 
\zeta \varepsilon \eta} \nabla^{\eta}H_{\gamma \mu}{}^{\zeta} 
\nabla^{\mu}H_{\alpha \beta \delta} + \frac{19}{32} H^{\alpha 
\beta \gamma} H^{\delta \epsilon \varepsilon} R_{\gamma \zeta 
\mu \eta} \nabla^{\eta}H_{\epsilon \varepsilon}{}^{\zeta} 
\nabla^{\mu}H_{\alpha \beta \delta}\nn\\&& + \frac{37}{48} H^{\alpha 
\beta \gamma} H^{\delta \epsilon \varepsilon} R_{\gamma \eta 
\mu \zeta} \nabla^{\eta}H_{\epsilon \varepsilon}{}^{\zeta} 
\nabla^{\mu}H_{\alpha \beta \delta} + \frac{85}{48} H^{\alpha 
\beta \gamma} H^{\delta \epsilon \varepsilon} R_{\gamma \zeta 
\varepsilon \eta} \nabla^{\eta}H_{\epsilon \mu}{}^{\zeta} 
\nabla^{\mu}H_{\alpha \beta \delta}\nn\\&& -  \frac{403}{48} 
H^{\alpha \beta \gamma} H^{\delta \epsilon \varepsilon} 
R_{\gamma \eta \varepsilon \zeta} \nabla^{\eta}H_{\epsilon 
\mu}{}^{\zeta} \nabla^{\mu}H_{\alpha \beta \delta} + 7 
H^{\alpha \beta \gamma} H^{\delta \epsilon \varepsilon} 
R_{\epsilon \zeta \varepsilon \eta} \nabla_{\mu}H_{\gamma}{}^{
\zeta \eta} \nabla^{\mu}H_{\alpha \beta \delta}\nn\\&& + 
\frac{5}{3} H^{\alpha \beta \gamma} H^{\delta \epsilon 
\varepsilon} R_{\gamma \zeta \varepsilon \eta} \nabla_{\mu}H_{
\epsilon}{}^{\zeta \eta} \nabla^{\mu}H_{\alpha \beta \delta} 
+ \frac{195}{8} H_{\alpha}{}^{\delta \epsilon} H^{\alpha \beta 
\gamma} R_{\epsilon \zeta \mu \eta} 
\nabla_{\varepsilon}H_{\delta}{}^{\zeta \eta} 
\nabla^{\mu}H_{\beta \gamma}{}^{\varepsilon}\nn\\&& + \frac{75}{8} 
H_{\alpha}{}^{\delta \epsilon} H^{\alpha \beta \gamma} 
R_{\varepsilon \zeta \mu \eta} \nabla^{\eta}H_{\delta 
\epsilon}{}^{\zeta} \nabla^{\mu}H_{\beta 
\gamma}{}^{\varepsilon} -  \frac{65}{8} H_{\alpha}{}^{\delta 
\epsilon} H^{\alpha \beta \gamma} R_{\varepsilon \eta \mu 
\zeta} \nabla^{\eta}H_{\delta \epsilon}{}^{\zeta} 
\nabla^{\mu}H_{\beta \gamma}{}^{\varepsilon}\nn\\&& + \frac{613}{48} 
H_{\alpha}{}^{\delta \epsilon} H^{\alpha \beta \gamma} 
R_{\epsilon \zeta \mu \eta} \nabla^{\eta}H_{\delta 
\varepsilon}{}^{\zeta} \nabla^{\mu}H_{\beta 
\gamma}{}^{\varepsilon} -  \frac{85}{24} H_{\alpha}{}^{\delta 
\epsilon} H^{\alpha \beta \gamma} R_{\epsilon \eta \mu \zeta} 
\nabla^{\eta}H_{\delta \varepsilon}{}^{\zeta} 
\nabla^{\mu}H_{\beta \gamma}{}^{\varepsilon}\nn\\&& -  \frac{103}{8} 
H_{\alpha}{}^{\delta \epsilon} H^{\alpha \beta \gamma} 
R_{\epsilon \zeta \varepsilon \eta} \nabla^{\eta}H_{\delta 
\mu}{}^{\zeta} \nabla^{\mu}H_{\beta \gamma}{}^{\varepsilon} + 
\frac{101}{12} H_{\alpha}{}^{\delta \epsilon} H^{\alpha \beta 
\gamma} R_{\epsilon \eta \varepsilon \zeta} 
\nabla^{\eta}H_{\delta \mu}{}^{\zeta} \nabla^{\mu}H_{\beta 
\gamma}{}^{\varepsilon} \nn\\&&-  \frac{451}{24} H_{\alpha}{}^{\delta 
\epsilon} H^{\alpha \beta \gamma} R_{\delta \zeta \epsilon 
\eta} \nabla^{\eta}H_{\varepsilon \mu}{}^{\zeta} 
\nabla^{\mu}H_{\beta \gamma}{}^{\varepsilon} -  \frac{79}{4} 
H_{\alpha}{}^{\delta \epsilon} H^{\alpha \beta \gamma} 
R_{\epsilon \zeta \varepsilon \eta} \nabla_{\mu}H_{\delta}{}^{
\zeta \eta} \nabla^{\mu}H_{\beta \gamma}{}^{\varepsilon} \nn\\&&+ 
\frac{19}{4} H_{\alpha}{}^{\delta \epsilon} H^{\alpha \beta 
\gamma} R_{\delta \zeta \epsilon \eta} 
\nabla_{\mu}H_{\varepsilon}{}^{\zeta \eta} 
\nabla^{\mu}H_{\beta \gamma}{}^{\varepsilon} - 31 
H_{\alpha}{}^{\delta \epsilon} H^{\alpha \beta \gamma} 
R_{\epsilon \zeta \mu \eta} \nabla_{\varepsilon}H_{\gamma}{}^{
\zeta \eta} \nabla^{\mu}H_{\beta \delta}{}^{\varepsilon}\nn\\&& -  
\frac{445}{48} H_{\alpha}{}^{\delta \epsilon} H^{\alpha \beta 
\gamma} R_{\varepsilon \zeta \mu \eta} \nabla^{\eta}H_{\gamma 
\epsilon}{}^{\zeta} \nabla^{\mu}H_{\beta 
\delta}{}^{\varepsilon} + \frac{469}{48} H_{\alpha}{}^{\delta 
\epsilon} H^{\alpha \beta \gamma} R_{\varepsilon \eta \mu 
\zeta} \nabla^{\eta}H_{\gamma \epsilon}{}^{\zeta} 
\nabla^{\mu}H_{\beta \delta}{}^{\varepsilon}\nn\\&& -  \frac{69}{2} 
H_{\alpha}{}^{\delta \epsilon} H^{\alpha \beta \gamma} 
R_{\epsilon \zeta \mu \eta} \nabla^{\eta}H_{\gamma 
\varepsilon}{}^{\zeta} \nabla^{\mu}H_{\beta 
\delta}{}^{\varepsilon} + \frac{65}{2} H_{\alpha}{}^{\delta 
\epsilon} H^{\alpha \beta \gamma} R_{\epsilon \eta \mu \zeta} 
\nabla^{\eta}H_{\gamma \varepsilon}{}^{\zeta} 
\nabla^{\mu}H_{\beta \delta}{}^{\varepsilon} \nn\\&&+ \frac{53}{3} 
H_{\alpha}{}^{\delta \epsilon} H^{\alpha \beta \gamma} 
R_{\epsilon \zeta \varepsilon \eta} \nabla^{\eta}H_{\gamma 
\mu}{}^{\zeta} \nabla^{\mu}H_{\beta \delta}{}^{\varepsilon} -  
\frac{47}{3} H_{\alpha}{}^{\delta \epsilon} H^{\alpha \beta 
\gamma} R_{\epsilon \eta \varepsilon \zeta} 
\nabla^{\eta}H_{\gamma \mu}{}^{\zeta} \nabla^{\mu}H_{\beta 
\delta}{}^{\varepsilon}\nn\\&& + \frac{545}{24} H_{\alpha}{}^{\delta 
\epsilon} H^{\alpha \beta \gamma} R_{\gamma \zeta \epsilon 
\eta} \nabla^{\eta}H_{\varepsilon \mu}{}^{\zeta} 
\nabla^{\mu}H_{\beta \delta}{}^{\varepsilon} + \frac{85}{6} 
H_{\alpha}{}^{\delta \epsilon} H^{\alpha \beta \gamma} 
R_{\epsilon \zeta \varepsilon \eta} \nabla_{\mu}H_{\gamma}{}^{
\zeta \eta} \nabla^{\mu}H_{\beta \delta}{}^{\varepsilon}\nn\\&& + H_{
\alpha}{}^{\delta \epsilon} H^{\alpha \beta \gamma} R_{\gamma 
\zeta \epsilon \eta} \nabla_{\mu}H_{\varepsilon}{}^{\zeta 
\eta} \nabla^{\mu}H_{\beta \delta}{}^{\varepsilon} -  
\frac{143}{48} H_{\alpha \beta}{}^{\delta} H^{\alpha \beta 
\gamma} R_{\varepsilon \zeta \mu \eta} \nabla^{\eta}H_{\delta 
\epsilon}{}^{\zeta} \nabla^{\mu}H_{\gamma}{}^{\epsilon 
\varepsilon}\nn\\&& + \frac{35}{48} H_{\alpha \beta}{}^{\delta} 
H^{\alpha \beta \gamma} R_{\varepsilon \eta \mu \zeta} 
\nabla^{\eta}H_{\delta \epsilon}{}^{\zeta} 
\nabla^{\mu}H_{\gamma}{}^{\epsilon \varepsilon} + 
\frac{275}{24} H_{\alpha \beta}{}^{\delta} H^{\alpha \beta 
\gamma} R_{\epsilon \zeta \varepsilon \eta} 
\nabla^{\eta}H_{\delta \mu}{}^{\zeta} 
\nabla^{\mu}H_{\gamma}{}^{\epsilon \varepsilon}\nn\\&& + \frac{11}{4} 
H_{\alpha \beta}{}^{\delta} H^{\alpha \beta \gamma} R_{\delta 
\zeta \mu \eta} \nabla^{\eta}H_{\epsilon 
\varepsilon}{}^{\zeta} \nabla^{\mu}H_{\gamma}{}^{\epsilon 
\varepsilon} -  \frac{17}{8} H_{\alpha \beta}{}^{\delta} 
H^{\alpha \beta \gamma} R_{\delta \eta \mu \zeta} 
\nabla^{\eta}H_{\epsilon \varepsilon}{}^{\zeta} 
\nabla^{\mu}H_{\gamma}{}^{\epsilon \varepsilon}\nn\\&& -  
\frac{35}{48} H_{\alpha \beta}{}^{\delta} H^{\alpha \beta 
\gamma} R_{\delta \zeta \varepsilon \eta} 
\nabla^{\eta}H_{\epsilon \mu}{}^{\zeta} 
\nabla^{\mu}H_{\gamma}{}^{\epsilon \varepsilon} + \frac{3}{4} 
H_{\alpha \beta}{}^{\delta} H^{\alpha \beta \gamma} R_{\delta 
\eta \varepsilon \zeta} \nabla^{\eta}H_{\epsilon 
\mu}{}^{\zeta} \nabla^{\mu}H_{\gamma}{}^{\epsilon \varepsilon} 
\nn\\&&+ \frac{209}{32} H_{\alpha \beta}{}^{\delta} H^{\alpha \beta 
\gamma} R_{\epsilon \zeta \varepsilon \eta} 
\nabla_{\mu}H_{\delta}{}^{\zeta \eta} 
\nabla^{\mu}H_{\gamma}{}^{\epsilon \varepsilon} + 5 H_{\alpha 
\beta}{}^{\delta} H^{\alpha \beta \gamma} R_{\delta \zeta 
\varepsilon \eta} \nabla_{\mu}H_{\epsilon}{}^{\zeta \eta} 
\nabla^{\mu}H_{\gamma}{}^{\epsilon \varepsilon}\labell{RPH2H2}
\eeqa
The corresponding Lagrangian in the scheme used in \cite{Garousi:2020gio} has  91 couplings.

Having set all parameters in the structures $H^8$, $RH^6$, $R^2H^4$,  $R^3H^2$,  $(\nabla H)^2H^4$  and $R(\nabla H)^2H^2$ to  zero, except the couplings in \reef{H8}, \reef{RH6}, \reef{R2H4}, \reef{R3H2}, \reef{PH2H4} and \reef{RPH2H2}, we now find the minimum number of couplings   with structure $R^2(\nabla H)^2$. To this end, we  again keep only the terms that if one set to zero any of them, the equation \reef{DLK} has no solution. In this case there are 22 terms. Setting all parameters in the structure $R^2(\nabla H)^2$ to zero, except those 22 parameters,  and solving \reef{DLK}, one finds the following couplings in the structure $R^2(\nabla H)^2$: 
\beqa
L_3^{R^2(\prt H)^2}&=&- \frac{5}{24} R_{\epsilon \mu \varepsilon \zeta} R^{\epsilon 
\varepsilon \mu \zeta} \nabla_{\delta}H_{\alpha \beta \gamma} 
\nabla^{\delta}H^{\alpha \beta \gamma} -  \frac{9}{2} 
R_{\gamma}{}^{\varepsilon \mu \zeta} R_{\epsilon \mu 
\varepsilon \zeta} \nabla_{\delta}H_{\alpha 
\beta}{}^{\epsilon} \nabla^{\delta}H^{\alpha \beta \gamma}\nn\\&& - 9 
R_{\beta}{}^{\mu}{}_{\epsilon}{}^{\zeta} R_{\gamma \zeta 
\varepsilon \mu} \nabla_{\delta}H_{\alpha}{}^{\epsilon 
\varepsilon} \nabla^{\delta}H^{\alpha \beta \gamma} + 
\frac{29}{4} R_{\beta}{}^{\mu}{}_{\epsilon}{}^{\zeta} 
R_{\gamma \mu \varepsilon \zeta} \nabla_{\delta}H_{\alpha}{}^{
\epsilon \varepsilon} \nabla^{\delta}H^{\alpha \beta \gamma}\nn\\&& + 
\frac{7}{4} R_{\beta}{}^{\mu}{}_{\gamma}{}^{\zeta} R_{\epsilon 
\mu \varepsilon \zeta} \nabla_{\delta}H_{\alpha}{}^{\epsilon 
\varepsilon} \nabla^{\delta}H^{\alpha \beta \gamma} + 
\frac{19}{2} R_{\gamma}{}^{\varepsilon \mu \zeta} R_{\epsilon 
\mu \varepsilon \zeta} \nabla^{\delta}H^{\alpha \beta \gamma} 
\nabla^{\epsilon}H_{\alpha \beta \delta}\nn\\&& -  \frac{5}{2} 
R_{\gamma}{}^{\mu}{}_{\epsilon}{}^{\zeta} R_{\delta \zeta 
\varepsilon \mu} \nabla^{\delta}H^{\alpha \beta \gamma} 
\nabla^{\varepsilon}H_{\alpha \beta}{}^{\epsilon} + 6 
R_{\gamma}{}^{\mu}{}_{\epsilon}{}^{\zeta} R_{\delta \mu 
\varepsilon \zeta} \nabla^{\delta}H^{\alpha \beta \gamma} 
\nabla^{\varepsilon}H_{\alpha \beta}{}^{\epsilon}\nn\\&& -  
\frac{5}{2} R_{\gamma}{}^{\mu}{}_{\delta}{}^{\zeta} 
R_{\epsilon \zeta \varepsilon \mu} \nabla^{\delta}H^{\alpha 
\beta \gamma} \nabla^{\varepsilon}H_{\alpha 
\beta}{}^{\epsilon} - 2 
R_{\beta}{}^{\mu}{}_{\epsilon}{}^{\zeta} R_{\gamma \zeta 
\varepsilon \mu} \nabla^{\delta}H^{\alpha \beta \gamma} 
\nabla^{\varepsilon}H_{\alpha \delta}{}^{\epsilon}\nn\\&& + 
\frac{43}{2} R_{\beta}{}^{\mu}{}_{\epsilon}{}^{\zeta} 
R_{\gamma \mu \varepsilon \zeta} \nabla^{\delta}H^{\alpha 
\beta \gamma} \nabla^{\varepsilon}H_{\alpha 
\delta}{}^{\epsilon} -  \frac{29}{2} 
R_{\beta}{}^{\mu}{}_{\gamma}{}^{\zeta} R_{\epsilon \mu 
\varepsilon \zeta} \nabla^{\delta}H^{\alpha \beta \gamma} 
\nabla^{\varepsilon}H_{\alpha \delta}{}^{\epsilon}\nn\\&& -  
\frac{3}{4} R_{\alpha \epsilon \beta \varepsilon} R_{\gamma 
\zeta \delta \mu} \nabla^{\delta}H^{\alpha \beta \gamma} 
\nabla^{\zeta}H^{\epsilon \varepsilon \mu} -  \frac{3}{8} 
R_{\alpha \beta \delta \epsilon} R_{\gamma \zeta \varepsilon 
\mu} \nabla^{\delta}H^{\alpha \beta \gamma} \nabla^{\zeta}H^{
\epsilon \varepsilon \mu}\nn\\&& -  \frac{29}{4} R_{\alpha \epsilon 
\beta \varepsilon} R_{\gamma \mu \delta \zeta} 
\nabla^{\delta}H^{\alpha \beta \gamma} 
\nabla^{\zeta}H^{\epsilon \varepsilon \mu} - 7 R_{\beta \mu 
\epsilon}{}^{\zeta} R_{\gamma \zeta \delta \varepsilon} 
\nabla^{\delta}H^{\alpha \beta \gamma} 
\nabla^{\mu}H_{\alpha}{}^{\epsilon \varepsilon} \nn\\&&- 8 R_{\beta 
\gamma \epsilon}{}^{\zeta} R_{\delta \zeta \varepsilon \mu} 
\nabla^{\delta}H^{\alpha \beta \gamma} 
\nabla^{\mu}H_{\alpha}{}^{\epsilon \varepsilon} + 18 R_{\beta 
\epsilon \gamma}{}^{\zeta} R_{\delta \mu \varepsilon \zeta} 
\nabla^{\delta}H^{\alpha \beta \gamma} 
\nabla^{\mu}H_{\alpha}{}^{\epsilon \varepsilon}\nn\\&& - 9 R_{\beta 
\delta \gamma}{}^{\zeta} R_{\epsilon \mu \varepsilon \zeta} 
\nabla^{\delta}H^{\alpha \beta \gamma} 
\nabla^{\mu}H_{\alpha}{}^{\epsilon \varepsilon} -  
\frac{21}{4} R_{\alpha \mu \beta}{}^{\zeta} R_{\gamma \zeta 
\epsilon \varepsilon} \nabla^{\delta}H^{\alpha \beta \gamma} 
\nabla^{\mu}H_{\delta}{}^{\epsilon \varepsilon}\nn\\&& -  
\frac{25}{2} R_{\alpha \epsilon \beta}{}^{\zeta} R_{\gamma 
\zeta \varepsilon \mu} \nabla^{\delta}H^{\alpha \beta \gamma} 
\nabla^{\mu}H_{\delta}{}^{\epsilon \varepsilon} + \frac{29}{2} 
R_{\alpha \epsilon \beta}{}^{\zeta} R_{\gamma \mu \varepsilon 
\zeta} \nabla^{\delta}H^{\alpha \beta \gamma} \nabla^{\mu}H_{
\delta}{}^{\epsilon \varepsilon}\labell{R2PH2}
\eeqa
The corresponding Lagrangian in the scheme used in \cite{Garousi:2020gio} has  also the same couplings. However, the coefficients of the couplings in the two schemes are all different. This is as expected, because if their coefficients were the same then the S-matrix of three-graviton-two-B-field dictates that the couplings in the structure  $R^3H^2$ in the present scheme, \ie couplings \reef{R3H2}, would be the same as the corresponding couplings in the scheme used in \cite{Garousi:2020gio}, which is not. 

Finally, having set all parameters in the structures $H^8$, $RH^6$, $R^2H^4$,  $R^3H^2$,  $(\nabla H)^2H^4$, $R(\nabla H)^2H^2$ and $R^2(\nabla H)^2$ to  zero, except the couplings in \reef{H8}, \reef{RH6}, \reef{R2H4}, \reef{R3H2}, \reef{PH2H4}, \reef{RPH2H2} and \reef{R2PH2}, we  find the minimum number of couplings   with structure $(\nabla H)^4$. To this end, we  again keep only the terms that if one set to zero any of them, the equation \reef{DLK} has no solution. In this case there are 12 terms. Setting all parameters in the structure $(\nabla H)^4$ to zero, except those 12 parameters,  and solving \reef{DLK}, one finds the following couplings in the structure $(\nabla H)^4$: 
\beqa
L_3^{(\prt H)^4}&=&\frac{1}{8} \nabla^{\delta}H^{\alpha \beta \gamma} 
\nabla_{\epsilon}H_{\gamma}{}^{\mu \zeta} 
\nabla_{\varepsilon}H_{\delta \mu \zeta} 
\nabla^{\varepsilon}H_{\alpha \beta}{}^{\epsilon} + 
\frac{281}{48} \nabla_{\delta}H_{\alpha}{}^{\epsilon 
\varepsilon} \nabla^{\delta}H^{\alpha \beta \gamma} 
\nabla_{\zeta}H_{\gamma \varepsilon \mu} 
\nabla^{\zeta}H_{\beta \epsilon}{}^{\mu} \nn\\&&-  \frac{275}{48} 
\nabla^{\delta}H^{\alpha \beta \gamma} \nabla^{\varepsilon}H_{
\alpha \beta}{}^{\epsilon} \nabla_{\zeta}H_{\epsilon 
\varepsilon \mu} \nabla^{\zeta}H_{\gamma \delta}{}^{\mu} -  
\frac{971}{96} \nabla^{\delta}H^{\alpha \beta \gamma} 
\nabla^{\varepsilon}H_{\alpha \beta}{}^{\epsilon} 
\nabla_{\zeta}H_{\delta \varepsilon \mu} 
\nabla^{\zeta}H_{\gamma \epsilon}{}^{\mu} \nn\\&& + \frac{397}{48} 
\nabla^{\delta}H^{\alpha \beta \gamma} 
\nabla_{\epsilon}H_{\delta \mu \zeta} 
\nabla^{\varepsilon}H_{\alpha \beta}{}^{\epsilon} 
\nabla^{\zeta}H_{\gamma \varepsilon}{}^{\mu} + \frac{279}{16} 
\nabla^{\delta}H^{\alpha \beta \gamma} \nabla^{\varepsilon}H_{
\alpha \beta}{}^{\epsilon} \nabla_{\zeta}H_{\delta \epsilon 
\mu} \nabla^{\zeta}H_{\gamma \varepsilon}{}^{\mu} \nn\\&& + 
\frac{275}{96} \nabla_{\delta}H_{\alpha \beta}{}^{\epsilon} 
\nabla^{\delta}H^{\alpha \beta \gamma} 
\nabla_{\zeta}H_{\epsilon \varepsilon \mu} 
\nabla^{\zeta}H_{\gamma}{}^{\varepsilon \mu} -  
\frac{397}{288} \nabla^{\delta}H^{\alpha \beta \gamma} 
\nabla^{\epsilon}H_{\alpha \beta \gamma} 
\nabla_{\zeta}H_{\epsilon \varepsilon \mu} 
\nabla^{\zeta}H_{\delta}{}^{\varepsilon \mu}  \nn\\&&+ \frac{61}{48} 
\nabla_{\gamma}H_{\varepsilon \mu \zeta} 
\nabla^{\delta}H^{\alpha \beta \gamma} 
\nabla^{\epsilon}H_{\alpha \beta \delta} 
\nabla^{\zeta}H_{\epsilon}{}^{\varepsilon \mu} -  
\frac{95}{24} \nabla^{\delta}H^{\alpha \beta \gamma} \nabla^{
\varepsilon}H_{\alpha \beta}{}^{\epsilon} 
\nabla^{\zeta}H_{\gamma \varepsilon}{}^{\mu} 
\nabla_{\mu}H_{\delta \epsilon \zeta} \nn\\&& + \frac{133}{96} 
\nabla^{\delta}H^{\alpha \beta \gamma} \nabla^{\varepsilon}H_{
\alpha \beta}{}^{\epsilon} \nabla^{\zeta}H_{\gamma 
\epsilon}{}^{\mu} \nabla_{\mu}H_{\delta \varepsilon \zeta} + 
\frac{69}{16} \nabla^{\delta}H^{\alpha \beta \gamma} \nabla^{
\varepsilon}H_{\alpha \beta}{}^{\epsilon} 
\nabla^{\zeta}H_{\gamma \delta}{}^{\mu} 
\nabla_{\mu}H_{\epsilon \varepsilon \zeta}\labell{PH4}
\eeqa
The corresponding Lagrangian in the scheme used in \cite{Garousi:2020gio} has  8 couplings. 

The couplings in \reef{RRf}, \reef{H8}, \reef{RH6}, \reef{R2H4}, \reef{R3H2}, \reef{PH2H4}, \reef{RPH2H2}, \reef{R2PH2} and \reef{PH4} are all NS-NS couplings at order $\alpha'^3$ in one particular scheme which has 251 couplings. We have found this number of couplings by first finding the minimum number of couplings with structure $H^8$, then the minimum number of couplings in the structure $RH^6$ and so on. If one changes the order of structures, one would find different number of couplings. For example, if one first find the minimal couplings in structure $R^3H^2$, then the minimal couplings in the structures $R^2H^4$, $RH^6$, $H^8$, $R(\nabla H)^2H^2$, $(\nabla H)^2H^4$, $R^2(\nabla H)^2$ and $(\nabla H)^4$, one would find that the minimal couplings in these sort of structures are $19, 7, 1, 4, 96, 95, 22, 13$, respectively. Adding these numbers to the 2 couplings in structure $R^4$ which is scheme independent, one finds 259 couplings in this scheme.  The minimal  couplings  in the structures $H^8$, $(\nabla H)^2H^4$, $R(\nabla H)^2H^2$, $R^3H^2$, $RH^6$, $R^2H^4$, $R^2(\nabla H)^2$,     $(\nabla H)^4$, $R^4$, are $2,45,87,21,28,49,22,14,2$, respectively, which add up to 270 couplings.   Even though we could not prove that the  251 couplings that we have found in this paper are the minimum number of couplings at this order, we have checked many other schemes which have only $R_{\mu\nu\alpha\beta}, H_{\mu\nu\alpha}, \nabla_{\mu}H_{\nu\alpha\beta}$. They all have more couplings than 251. 

We have found that there are  schemes in which the dilaton appears only as the overall factor $e^{-2\Phi}$ in the NS-NS couplings in the string frame. In the Einstein frame $g_{\mu\nu}\stackrel{E}{\longrightarrow}e^{\Phi/2}g_{\mu\nu}$, however,  the derivative of dilaton appears through the following replacements: 
\beqa
g^{\mu\nu}&\stackrel{E}{\longrightarrow}&e^{-\Phi/2}g^{\mu\nu}\\
R_{\mu\nu\alpha\beta}&\stackrel{E}{\longrightarrow}&e^{\Phi/2}R_{\mu\nu\alpha\beta}+\frac{e^{\Phi/2}}{16}(g_{\alpha\nu}g_{\beta\mu}-g_{\alpha\mu}g_{\beta\nu})\nabla_\gamma\Phi\nabla^\gamma\Phi\nn\\
&&+\frac{e^{\Phi/2}}{16}(g_{\alpha\mu}\nabla_\beta\Phi\nabla_\nu\Phi -g_{\alpha\nu}\nabla_\beta\Phi\nabla_\mu\Phi+g_{\beta\nu}\nabla_\alpha\Phi\nabla_\mu\Phi-g_{\beta\mu}\nabla_\alpha\Phi\nabla_\nu\Phi)\nn\\
&&-\frac{e^{\Phi/2}}{4}(g_{\alpha\mu}\nabla_\beta\nabla_\nu\Phi-g_{\alpha\nu}\nabla_\beta\nabla_\mu\Phi+g_{\beta\nu}\nabla_\alpha\nabla_\mu\Phi-g_{\beta\mu}\nabla_\alpha\nabla_\nu\Phi)\nn\\
\nabla_\mu H_{\nu\alpha\beta}&\stackrel{E}{\longrightarrow}&\nabla_\mu H_{\nu\alpha\beta}+\frac{1}{4}(g_{\alpha\mu}\nabla^\gamma\Phi H_{\gamma\beta\nu}-g_{\beta\mu}\nabla^\gamma\Phi H_{\gamma\alpha\nu}+g_{\mu\nu}\nabla^\gamma\Phi H_{\gamma\alpha\beta})\nn\\
&&-\frac{3}{4}\nabla_\mu\Phi H_{\nu\alpha\beta}+\frac{1}{4}(\nabla_\alpha\Phi H_{\beta\mu\nu}-\nabla_\beta\Phi H_{\alpha\mu\nu}-\nabla_\nu\Phi H_{\alpha\beta\mu})\nn
\eeqa
The leading order effective action \reef{S0bf} in the Einstein frame becomes
\beqa
 \bS_0^E= -\frac{2}{\kappa^2}\int d^{10} x\sqrt{-g} \Big[ R - \frac{1}{2}\nabla_{a}\Phi \nabla^{a}\Phi-\frac{e^{-\Phi}}{12}H^2+ \sum_{n=1}^{9}\frac{e^{\frac{5-n}{2}\Phi}}{n!}F^{(n)}\cdot F^{(n)}\Big]\labell{ES0bf}
\eeqa
Similarly one can transform the 251 $\alpha'^3$-order couplings in \reef{S3bf1} to the Einstein frame. Then one can confirm the resulting couplings for metric, B-field and dilaton with the corresponding sphere-level S-matrix elements of NS-NS vertex operators in the superstring theory.

The  couplings \reef{PH4}, \reef{R2PH2} and \reef{RRf} have at least four fields in the Einstein frame. Hence, it is easy to check them with the corresponding string theory S-matrix element of four NS-NS vertex operators which has only contact terms at eight-momentum level. This check has been done in \cite{Garousi:2020gio} for the corresponding couplings in the scheme used in \cite{Garousi:2020gio} in the string frame.
The S-matrix element of four NS-NS vertex operators in the superstring theory has been calculated in \cite{Schwarz:1982jn,Gross:1986iv}. The low energy expansion of this S-matrix element produces the following eight-derivative couplings in the Einstein frame \cite{Gross:1986mw,Myers:1987qx,Policastro:2008hg,Paban:1998ea,Garousi:2013zca}:
\beqa
\bS_3^E&\supset& \frac{\gamma_3}{\kappa^2}\int d^{10}x\sqrt{-g} e^{-3\Phi/2} \cL(\cR) \labell{S2}
 \eeqa  
 where the normalization factor is $\gamma_3=\z(3)/2^5$, and the Lagrangian density has the following eight independent terms:
 \beqa
 \cL(\cR)&=&  \frac{1}{8}\cR_{\kappa \gamma}{}^{\delta \beta} \cR^{\kappa \gamma \tau \alpha}\cR_{\mu \nu \delta \tau} \cR^{\mu \nu}{}_{\alpha \beta} + \frac{1}{32}\cR_{\alpha \beta \kappa \gamma} \cR^{\alpha \beta \kappa \gamma}\cR_{\mu \nu \delta \tau} \cR^{\mu \nu \delta \tau}\nn\\&& + \frac{1}{16}\cR_{\alpha \beta}{}^{\delta \tau} \cR^{\alpha \beta \kappa \gamma}\cR_{\mu \nu \delta \tau} \cR^{\mu \nu}{}_{\kappa \gamma} -  \frac{1}{4}\cR_{\kappa \gamma}{}^{\alpha \beta} \cR^{\kappa \gamma \delta}{}_{\alpha}\cR_{\mu \nu \delta \tau} \cR^{\mu \nu \tau}{}_{\beta}\nn\\&& + \frac{1}{4} \cR^{\alpha}{}_{\beta \kappa \gamma}\cR_{\mu \nu \delta \tau} \cR^{\mu \beta \kappa \gamma} \cR^{\nu}{}_{\alpha}{}^{\delta \tau} + \frac{1}{8} \cR^{\alpha}{}_{\beta}{}^{\delta \tau}\cR_{\mu \nu \delta \tau} \cR^{\mu \beta \kappa \gamma} \cR^{\nu}{}_{\alpha \kappa \gamma}\nn\\&& + \frac{1}{2} \cR^{\alpha}{}_{\beta}{}^{\kappa}{}_{\gamma}\cR_{\mu \nu \delta \tau} \cR^{\mu \beta \delta \gamma} \cR^{\nu}{}_{\alpha}{}^{\tau}{}_{\kappa} + \cR^{\alpha}{}_{\beta}{}^{\delta}{}_{\gamma}\cR_{\mu \nu \delta \tau} \cR^{\mu \beta \kappa \gamma} \cR^{\nu}{}_{\alpha}{}^{\tau}{}_{\kappa}\labell{L}
 \eeqa
 where   $\cR_{\mu\nu\alpha\beta}$ is the linear part of the following tensor in which the flat metric is perturbed as  $g_{\mu\nu}=\eta_{\mu\nu}+h_{\mu\nu}$:
\beqa
\cR_{\mu\nu\alpha\beta}&=& R_{\mu\nu\alpha\beta}-\frac{1}{4}(g_{\alpha\mu}\nabla_\beta\nabla_\nu\Phi-g_{\alpha\nu}\nabla_\beta\nabla_\mu\Phi+g_{\beta\nu}\nabla_\alpha\nabla_\mu\Phi-g_{\beta\mu}\nabla_\alpha\nabla_\nu\Phi)+e^{-\Phi/2}H_{\mu\nu[\alpha;\beta]}\nn
\eeqa 
Here $H_{\mu\nu[\alpha;\beta]}=\frac{1}{2}\nabla_{\beta}H_{\mu\nu\alpha}-\frac{1}{2}\nabla_{\alpha}H_{\mu\nu\beta}$.
Now, using the perturbation  $g_{\mu\nu}=\eta_{\mu\nu}+h_{\mu\nu}$ for the metric in the couplings  in \reef{RRf}, \reef{R2PH2} and  \reef{PH4},  one can find four-field couplings in the Einstein frame. Then one should transform them to the momentum space and use the conservation of momentum, \ie $k^1+k^2+k^3+k^4=0$ where $k_i$ for $i=1,2,3,4$ is the momentum of the external state $i$.  Using the on-shell conditions $k_i\cdot k_i=0$, $ k_i\cdot \z_i=0$, $ k_i\cdot \theta_i=0$  where the graviton polarization is $\epsilon_i^{\mu\nu}=\z_i^{\mu}\z_i^{\nu}$,  the B-field polarization is $\epsilon_i^{\mu\nu}=\z_i^\mu\theta_i^\nu-\z_i^\nu\theta_i^\mu$ and the dilaton polarization is one, one finds that the resulting four-point functions are exactly the same as the four-point functions corresponding to the couplings \reef{L}. Hence, the couplings  \reef{RRf}, \reef{R2PH2} and  \reef{PH4} are consistent with the S-matrix element of four NS-NS vertex operators. 

One may compare the five-point functions constructed from the couplings in this paper with the corresponding sphere-level S-matrix element of five NS-NS vertex operators \cite{Liu:2019ses}. A crucial property of the sphere-level S-matrix elements in the superstring theory is that the S-matrix elements should not have terms with zero and one Mandelstam variables \cite{Garousi:2012yr,Barreiro:2012aw}. If the couplings we have found is going to reproduce the low energy expansion of the superstring theory S-matrix elements, then the S-matrix elements constructed from the couplings in \reef{ES0bf} and the couplings in the Einstein frame form of \reef{S3bf1} should have no term with zero and one Mandelstam variables. This would be a non-trivial  check for the couplings \reef{S3bf1}. This check has been done  for five-graviton S-matrix element in \cite{Garousi:2012yr}. It would be interesting to check this consistency for five-point and higher S-matrix elements involving B-field and dilatons.

When there is no B-field, in the Einstein frame, there are couplings in \reef{S3bf1} which involve odd number of dilaton. They are not consistent with S-duality of the type IIB superstring theory \cite{Garousi:2013qka}. It means one has to use field redefinition to remove the odd dilaton couplings. In fact the S-matrix element of one dilaton and three gravitions and the S-matrix element of three dilatons and one graviton are zero \cite{Garousi:2012yr}. This indicates that, by using field redefinitions, the corresponding couplings in the field theory \reef{S3bf1} in the Einstein frame  can be removed. In the new field variables, then one may include appropriate R-R couplings in the type IIB theory to make the effective action to be invariant  under the S-duality. In that way one may include some parts of the R-R couplings of type IIB theory in the effective action \reef{S3bf1}. If one ignores the dilaton and the R-R scalar couplings, then the S-duality dictates that the  couplings involving the metric and the R-R two-form  are exactly the same as the couplings  in \reef{S3bf1} in which one  replaces $H\rightarrow F$ where $F$ is the R-R two-form field strength. 

Another approach for including the R-R couplings in \reef{S3bf1} is to first find all independent gauge invariant couplings involving all NS-NS and R-R couplings, as has been done in \cite{Garousi:2020mqn} for the NS-NS couplings. Then one may use the T-duality constraint on the couplings to find the coeffients of the gauge invariant couplings, as has been done in \cite{Garousi:2020gio} for the NS-NS couplings.  We leave the details of the calculations for including the R-R couplings in  \reef{S3bf1} to the future works.



\end{document}